\documentclass[fleqn,usenatbib]{mnras}

\usepackage{newtxtext,newtxmath}

\usepackage[T1]{fontenc}
\usepackage{ae,aecompl,pdflscape}

\usepackage{graphicx}	
\usepackage{amsmath}	
\usepackage{float}
\usepackage{epstopdf}
\usepackage{subfigure}
\usepackage{ulem}
\usepackage{array}
\usepackage{booktabs}
\usepackage{supertabular}
\usepackage{multicol} 

\DeclareRobustCommand{\VAN}[3]{#2}
\let\VANthebibliography\thebibliography
\def\thebibliography{\DeclareRobustCommand{\VAN}[3]{##3}\VANthebibliography}

\usepackage{graphicx}	

\usepackage{lipsum}

\title[AGN in Red Geysers]{Active Galactic Nuclei signatures in Red Geyser galaxies from Gemini GMOS-IFU observations}

\author[G. S. Ilha et al.]{
Gabriele S. Ilha,$^{1,2}$\thanks{E-mail:gabriele.ilha@acad.ufsm.br}
Rogemar A. Riffel,$^{1,2}$
Tiago V. Ricci,$^{3}$
Sandro B. Rembold,$^{1,2}$ 
\newauthor
Thaisa Storchi-Bergmann,$^{4,2}$
Rog\'erio Riffel,$^{4,2}$
Namrata Roy,$^{7}$
Kevin Bundy,$^{7,8}$
\newauthor
Rodrigo Nemmen,$^{5}$
J\'aderson S. Schimoia,$^{1,2}$
Luiz N. da Costa$^{2,6}$\\
$^{1}$Departamento de F\'isica, CCNE, Universidade Federal de Santa Maria, 97105-900, Santa Maria, RS, Brazil\\
$^{2}$Laborat\'orio Interinstitucional de e-Astronomia - LIneA, Rua Gal. Jos\'e Cristino 77, Rio de Janeiro, RJ - 20921-400, Brazil\\
$^{3}$Campus Cerro Largo, Universidade Federal da Fronteira Sul, RS 97900-000, Brazil\\
$^{4}$ Departamento de Astronomia, Universidade Federal do Rio Grande do Sul. 9500 Bento Gonçalves, Porto Alegre, 91501-970, Brazil\\
$^{5}$Instituto de Astronomia, Geof\'isica e Ci\^encias Atmosf\'ericas, Universidade de S\~ao Paulo, S\~ao Paulo SP 05508-090, Brazil\\
$^{6}$Observat\'orio Nacional - MCT, Rua General Jos\'e Cristino 77, Rio de Janeiro, RJ - 20921-400, Brazil \\
$^{7}$Department of Astronomy and Astrophysics, University of California, 1156 High Street, Santa Cruz, CA 95064, USA\\
$^{8}$UCO/Lick Observatory, Department of Astronomy and Astrophysics, University of California, 1156 High Street, Santa Cruz, CA 95064, USA\\
}

\date{Accepted XXX. Received YYY; in original form ZZZ}

\pubyear{2022}

\begin{document}
\label{firstpage}
\pagerange{\pageref{firstpage}--\pageref{lastpage}}
\maketitle

\begin{abstract}
Red Geysers are quiescent galaxies with galactic scale ionised outflows, likely due to low-luminosity Active Galactic
Nuclei (AGN). We used Gemini GMOS-IFU observations of the inner $\sim$1--3 kpc of nine Red Geysers selected from the MaNGA survey to study the gas ionisation and kinematics. The emission-line ratios suggest the presence of Seyfert/LINER (Low Ionisation Nuclear Emission Region) nuclei in all sources. Two galaxies show H$\alpha$ equivalent width (H$\alpha$ EW) larger than 3 \AA\ (indicative of AGN ionisation) within an aperture 2\farcs5 of diameter (1.3--3.7 kpc at the distance of galaxies) for MaNGA data, while with the higher resolution GMOS data, four galaxies present H$\alpha$ EW$>3$\,\AA\ within an aperture equal to the angular resolution (0.3--0.9 kpc). For two objects with GMOS-IFU data, the H$\alpha$ EW is lower than 3 \AA\ but larger than 1.5 \AA, most probably due to a faint AGN. The spatially resolved electron density maps show values between 100--3000 cm$^{-3}$ and are consistent with those determined in other studies. The large (MaNGA) and the nuclear scale (GMOS-IFU)  gas velocity fields are misaligned, with a kinematic position angle difference between 12$^{\circ}$ and 60$^{\circ}$. The [N\,{\sc ii}]$\lambda$6583 emission-line profiles are asymmetrical, with blue wings on the redshifted side of the velocity field and red wings on the blueshifted side. Our results support previous indications that the gas in  Red Geysers is ionised by an AGN, at least in their central region, with the presence of outflows, likely originating in a precessing accretion disc.
\end{abstract}



\section{Introduction}
\label{intro}
Galaxies in general present bi-modalities in several properties \citep{strateva01, blanton03, kauffmann03a, baldry04,  Bell04, mateus06, salim07,wetzel12}, such as colours \citep{strateva01,salim07}, D$_{n}(4000)$ index \citep{kauffmann03a}, and star formation rate (SFR)  \citep{mateus06,wetzel12}. Such bi-modalities reveal a dichotomy between two major classes (or ``sequences'') of galaxies: blue (star-forming, SF) and red (passive) objects. Blue sequence galaxies show ongoing star formation, young stellar populations and lower stellar masses. Meanwhile, red sequence galaxies have older stellar populations, higher stellar masses, and quenched star formation. Active Galactic Nuclei (AGN) hosts present colours and SFRs between those of the red and blue sequences \citep{mateus06,sanchez18}. A similar result was found by \citet{schawinski07}, but restricted to a sample of early-type galaxies: star-forming objects are in the blue sequence and objects without emission lines are in the red sequence (passive galaxies). Composite galaxies (star-forming+AGN) lie between these two regions, while galaxies whose gas ionization source is dominated by an AGN (Seyfert and LINER) are close to/in the red sequence.  These results led to AGN feedback being considered a possible mechanism to quench star formation and make the galaxy evolve from star-forming to quiescent. However, the nature of the mechanisms leading to the transformation of galaxies from the blue to the red sequence is an open question in our understanding of galaxy evolution. 


 AGN feedback has been suggested as a possible explanation for the relation between the mass of the central supermassive black hole (SMBH) and the stellar velocity dispersion of the bulge, the M$_{BH}$--$\sigma$ relation \citep[e.g.][]{Magorrian98,ferrarese00,gebhardt00, kormendy13}. Furthermore, it is an important ingredient in cosmological simulations; if not considered, such simulations result in galaxies with larger stellar masses than observed  \citep{diMatteo05,springel05,bower06}. Thus, AGN feedback is invoked as a mechanism that can suppress star formation in the host galaxies \citep{canodiaz12, cresci2015, carniani16, wylezalek16}, regulating their growths.  Besides this ``negative feedback'', AGN can also induce star formation in a few cases, that are then considered ``positive feedback'' \citep{Ishibashi2013, cresci2015, Maiolino17, Mallmann18, gallagher19, rogerio21}.

Negative AGN feedback present two main modes of operation: {\it radio/maintenance mode} occurs when the jets from the SMBH accretion disc heat the gas of the host galaxy and circumgalactic medium, preventing it from cooling \citep{croton06} and maintaining a low star formation rate. Radio mode feedback occurs in AGN with luminosity (L) much smaller than the Eddington luminosity \citep[L$\rm \leq0.01$L$_{Edd}$;][]{fabian12}.

The second mode of negative feedback is the {\it radiative/wind mode}, that can quench star formation in the host galaxy \citep{diMatteo05}, as winds from the AGN remove gas from the galaxy's gravitational potential or heat a large amount of gas preventing the star formation. In this feedback mode, typically found in quasars, the AGN luminosity is close to the Eddington limit \citep{fabian12}.

\citet{cheung16}, using MaNGA (Mapping Nearby Galaxies at APO) integral-field spectroscopy (IFS) data from the Sloan Digital Sky Survey-IV \citep[SDSS-IV; ][]{bundy15, drory15,law15,yan16,wake17,blanton17}, found a new class of quiescent galaxies ($NUV-r>5$) that show narrow bi-symmetric features in the  H$\alpha$ equivalent width resolved maps aligned with the gradient of the gas velocity field and misaligned with stellar velocity field. These galaxies became known as Red Geysers.  They present gas velocity fields with an amplitude reaching $\pm 200-300$ km s$^{-1}$ and being at least 1.5 times larger than the amplitude of the stellar velocity field \citep{cheung16,roy18}. 

Studying in detail the Akira galaxy, the prototype of the Red Geyser class, \citet{cheung16} showed that it presents large scale ionised bipolar outflows probably driven by a low-luminosity AGN (LLAGN), suggesting that the large scale gas kinematics in Red Geysers is produced by AGN driven winds. Using optical long-slit spectroscopy for two Red Geysers, \citet{roy21a} found that the emission lines present strong asymmetries in locations along the bi-symmetric pattern of the H$\alpha$ equivalent width  map. The emission-line profiles show blue wings on the redshifted side of the velocity field and red wings on the blueshifted side, being interpreted as winds signatures \citep{roy21a}.  \citet{riffel19}, by combining large scale (MaNGA) and nuclear region IFS (GMOS--Gemini Multi-Object Spectrographs \citep{ smith02, hook04}), found that the ionised outflow observed for the Akira galaxy changes the orientation about 50$^{\circ}$ from the nucleus to kpc scales and it is consistent with winds originated in a precessing accretion disc.  

So far, studies on Red Geyser galaxies suggest the presence of low-luminosity AGN responsible for winds driven by the central source \citep{cheung16, roy18, roy21c}. AGN signatures have been found in Red Geysers, mainly based on radio observations, as observed in Akira \citep{cheung16}. Furthermore, \citet{roy18,roy21c} demonstrated that Red Geysers have an excess of radio emission when compared to non-active galaxies, and this was interpreted as due to the presence of low-luminosity radio-mode AGNs. 

The low angular resolution of the MaNGA data (2\farcs5) limits the detection of low-luminosity AGN using optical diagnostic diagrams, as the central aperture includes emission from gas that is located up to 1.3--3.7 kpc from the nucleus (on average). Higher spatial resolution data can be used to map the ionised gas emission down to the inner few hundred pc (300--900 pc) of Red Geysers and verify whether it is consistent with AGN ionisation or not. However, such observations -- with angular resolution about four times better than that of MaNGA -- are available so far only for two Red Geysers \citep{riffel19,roy21a}. In both cases, the  [N\,{\sc ii}]$\lambda$6583/H$\alpha$ and  [O\,{\sc iii}]$\lambda$6583/H$\beta$ line ratios are typical of low-ionisation nuclear emission-line regions (LINERs). However, only Akira shows  H$\alpha$ EW $>3$ \AA\ \citep{riffel19}, which is a signature of photoionisation by an LLAGN \citep{cid11}. Thus, increasing the number of higher resolution optical observations is essential to characterise the nature of the gas emission and of the outflows in Red Geysers.

The outflows seen in Red Geysers are observed to extend up to $\sim$ 5 kpc, more than five times the typical extent of the NLR (Narrow Line Region) of Seyfert galaxies \citep{fischer13}. Furthermore, these galaxies correspond to about 5-10\% of the quiescent populations with stellar masses around 10$^{10}$ M$_{\odot}$ \citep{cheung16}. Thus, they are an essential piece to understand low-luminosity AGN feedback and its role in keeping the galaxy quiescent. \citet{roy21c} analysed the radio morphology of a sample of 42 Red Geysers, showing that galaxies with extended radio emission have a lower star formation rate than those with a compact radio source. This result can be related to the {\it radio/maintenance mode} feedback effect in the interstellar medium.  Furthermore,
\citet{roy21b} estimated the cool neutral gas mass of $10^{7}$--$10^{8}$ M$_{\odot}$ in the central region of Red Geysers, which could trigger a star formation rate of $\sim$ 1.0 M$_{\odot}$ yr$^{-1}$, but these objects display a much lower SFR $\sim$ 0.01 M$_{\odot}$ yr$^{-1}$. The {\it radio/maintenance mode} feedback may explain the low SFR observed in these galaxies, once the radio jet can heat the gas preventing it from forming stars.

Red Geysers are quiescent galaxies that show outflows on galactic scales, so they are interesting for studying feedback effects. However, to fully characterise the feedback in these objects, it is necessary to verify the origin of the gas emission and of the outflows. We use Gemini GMOS IFS  to map the ionised gas emission in the nuclear region of nine Red Geysers at spatial resolutions of  0.3--0.9 kpc, and investigate the gas ionisation source using optical-emission line ratio diagnostic diagrams. We also present an analysis of the main kinematic properties of these objects. A detailed analysis of the ionised gas kinematics,
including the modelling and estimates of the outflow properties, will be presented in a forthcoming work (Ilha et al., in prep.).

In this work, we adopt the cosmological parameters $h =0.7$, $\Omega_{m}= 0.3$, $\Omega_{\Lambda} = 0.7$. This paper is organised as follows: Sec. \ref{sample} describes the MaNGA data, sample selection, observations and data reduction.Sec. \ref{sec:spectralfitting} presents the spectral fitting procedure. The results are shown in Sec. \ref{Sec:results} and discussed in Sec. \ref{discussion}, while Sec. \ref{conclusions} summarises our conclusions. 

\section{The sample, observations and data reduction}
\label{sample}
\subsection{The MaNGA data}

 The MaNGA survey obtained optical integral field spectra using the 2.5-meter Sloan telescope \citep{gunn06,smee13} for a sample of about 10000 nearby galaxies \citep{bundy15, drory15,law15,yan16,wake17,blanton17}. MaNGA data have spectral coverage in the range 3600--10400 \AA. The field of view (FoV) depends on how many fibres were used to observe the science object, which can vary between 19--127 fibres (covering a FoV from 12\farcs0 to 32\farcs0 in diameter). In the MaNGA survey, data cubes are provided by the DRP \citep[Data Reduction Pipeline;][]{law16} for each galaxy. DRP also provides the reconstructed point spread function (PSF) at the $griz$ photometric bands  for each MaNGA data cube. To estimate the PSF the DRP uses a numerical simulation with the speciﬁc observing conditions of each exposure \citep{law16}. The mean $g$-band PSF full width at half maximum (FWHM) for 1390 MaNGA galaxies included in the
SDSS Data Release 13  \citep[DR13;][]{Albareti17}  is $\sim$2\farcs54 \citep{law16}. For the 9 galaxies in our GMOS sample, the MaNGA data cubes have  PSF FWHM of $\sim$2\farcs52 at the $g$-band and 2\farcs50 at the $r$-band. Thus, we will adopt 2\farcs50 as the angular resolution of the MaNGA data. In addition, fluxes, velocities,  velocity dispersion, equivalent widths for the strongest emission lines and also stellar velocity and stellar velocity dispersion measurements, among other properties derived from data cubes are provided by the Data Analysis Pipeline \citep[DAP;][]{westfall19,belfiore19}. For this work, we have used DAP data products from the MaNGA Product Launch-8 (MPL-8). MPL-8 contains 6779 data cubes and DAP products for $\sim$6520 galaxies, which are now public through the SDSS DR17 \citep{Abdurrouf22}. 
DAP fits the stellar kinematics using the Penalised
Pixel-fitting ({\sc ppxf}) routine \citep{cappellari04,cappellari17}. In the MPL-8, the MILES-HC \citep{westfall19} stellar spectra library was used as template to fit the stellar continuum and also to determine the stellar kinematics. According to  \citet{westfall19}, MILES-HC was constructed dividing the MILES stellar library \citep{falcon11, sanchez06} into 49 groups. In each group, the mean stellar spectra was obtained, which leads to 49 template spectra. However, spectra with prominent emission lines and those with a low signal-to-noise ratio were excluded, thus 42 stellar spectra are used to represent the stellar population contribution. In addition, DAP also fits the emission-line profiles with single Gaussian curves \citep{belfiore19}. We used  SPX-MILESHC-
MILESHC data products,  which include an analysis of each individual spaxel (SPX) by DAP with MILES-HC to determine the stellar kinematics and to fit the stellar continuum.
 
\subsection{Selection of Red Geysers from MaNGA}
\label{selection}
 Red geysers are quiescent galaxies, originally identified by \citet{cheung16} to present rest-frame colour $NUV-r > 5$.  Furthermore, \citet{roy18} selected objects with low SFR, $\rm log(SFR[M_{\odot}$ yr$^{-1}$])$<-2$, to remove possible obscured star formation.
 Thus, only galaxies that present $NUV-r > 5$ and $ \rm log(SFR[M_{\odot}$ yr$^{-1}$])$<-2$ were included in our sample. The absolute magnitudes in the $NUV$ \citep[from GALEX;][]{Martin05}  and $r$ (from SDSS) bands were extracted from the NASA-Sloan Atlas\footnote{\url{http://www.nsatlas.org}} (NSA) catalogue. The values for star formation rate were obtained from \citet{chang15}, which combines photometric data from the SDSS  sample with the Wide-Field Infrared Survey Explorer (WISE) data to determine the spectral energy distributions (SEDs) and star formation rates for more than 800,000 galaxies. In addition, Red Geysers present the following properties, as described by \citet{cheung16} and \citet{roy18}: (i) 
 a bi-symmetric pattern in the resolved map of H$\alpha$ equivalent width, which was interpreted by \citet{cheung16} as enhanced due to shocks or density increase along the outflow axis;
(ii) a bi-symmetric pattern seen on the map of H$\alpha$ equivalent width aligned with the gradient of the gas velocity field
and misaligned with the stellar velocity field gradient; and
(iii) a gas velocity field reaching values of up to 250 km s$^{-1}$, misaligned with the stellar velocity field and with gas velocity amplitude at least 1.5 times the amplitude of the stellar velocity field. These considerably  high gas velocity values suggest that the kinematics of the Red Geysers cannot be explained only by motion due to the gravitational potential of the galaxy. The disturbed kinematics is likely due to ionised gas outflows \citep{cheung16,riffel19,roy21a}.

To quantify the misalignment between the gas and stellar velocity fields we calculated the difference between the global kinematic position angles PA obtained for H$\alpha$ and stellar velocity fields, as $\rm \Delta PA=|PA_{gas}-PA_{stellar}|$. The global kinematic position angle (PA), or the orientation of the line of nodes, was determined using the IDL routine {\sc fit kinematics pa}\footnote{This routine was developed by M. Cappellari and is available at
\url{http://www-astro.physics.ox.ac.uk/~mxc/software}}, which is an implementation of the method presented in \cite{krajnovic06}. The gas and stellar velocity fields were considered to be misaligned when $\rm 10^{\circ}< \Delta PA<80^{\circ}$ and $\rm 100^{\circ}< \Delta PA<170^{\circ}$, because these criteria remove co-rotating, polar and counter rotating discs from the sample.
To determine the ratio between the gas and stellar velocity amplitudes, we determine the velocity amplitude by calculating the mean velocity value for 10\% of the spaxels with the highest absolute velocity values.  After these steps, we visually checked if the Red Geyser candidate galaxy showed the bi-symmetric pattern on H$\alpha$ equivalent width map as described in \cite{roy18}.

We found 92 Red Geysers in MPL-8, which comprises $\sim$1.4\% of the galaxies with DAP data in this MPL. \cite{roy18} found 84 Red Geysers in MPL-5, which is approximately 3\% of the galaxies with DAP measurements available in MPL-5. \citet{roy21b} updated the sample of \cite{roy18} with the MPL-9 data, which contains $\sim$ 8080 DAP objects, 140 or 1.7\% of them being Red Geysers. \citet{sanchez20} found that Red Geysers comprise less than 4\% of the elliptical galaxies and less than 1\% of the total galaxy population. 
We follow the selection criteria defined in \citet{roy18}, but there are some differences in our selection process.  \citet{roy18} visually determined the misalignment between the stellar and gas velocity fields (which are aligned to the bi-symmetric EW feature), while we quantified these differences using $\Delta$PA. Furthermore, we compared the amplitude of the velocity fields (stellar and gas) using only the 10\% of the spaxels with the highest absolute velocity values, while this was done by visual inspection in \citet{roy18}. The spaxels with velocity uncertainties greater than 25 km s$^{-1}$ were excluded from these calculations. Thus, the different percentage of galaxies between \citet{roy18} and our sample may be due to the use of better defined quantities in our selection.

\begin{figure}
	\includegraphics[width=0.45\textwidth]{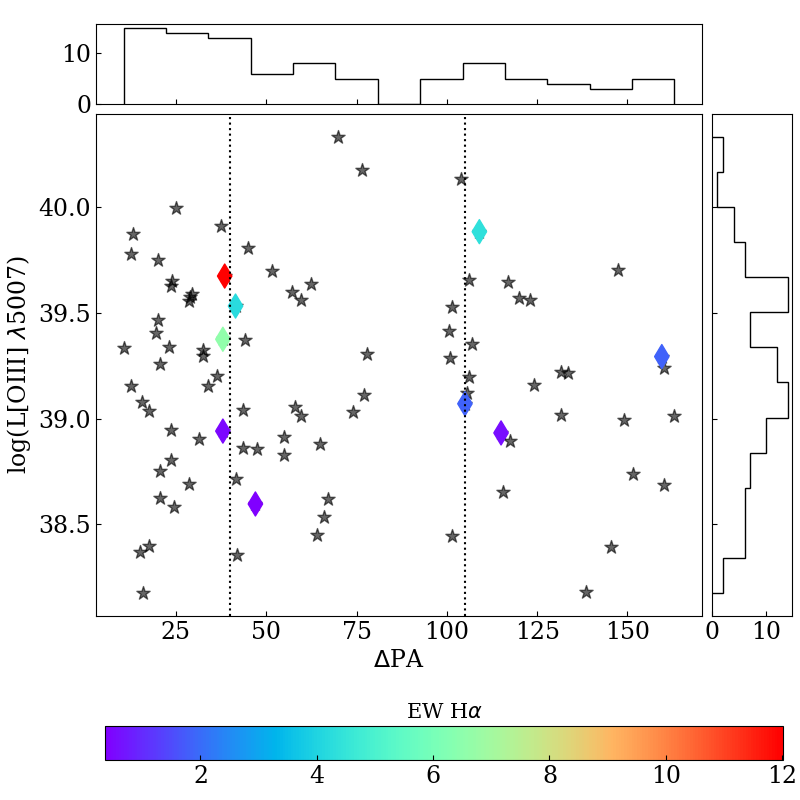}
    \caption{Plot of $\rm \Delta PA$ vs. [O\,{\sc iii}]$\lambda5007$ luminosity (L$_{\rm [OIII]}$) for the sample of Red Geysers selected from MaNGA (gray stars). Galaxies observed with GMOS-IFU are shown as thin diamond with colours according to the H$\alpha$ EW values of the colour bar  and the dotted vertical lines indicate the boundaries of the $\rm \Delta PA$  groups. The H$\alpha$ EW values were determined using GMOS-IFU data with an aperture equal to the angular resolution of Table \ref{exp}.}
    \label{fig:palum}
\end{figure}

\begin{table*}
\centering
\caption{Sample of Red Geysers observed with GMOS-IFU. (1) Galaxy identification in the MaNGA survey. (2) Redshift from the NSA catalogue. (3)  Rest-frame colour $NUV-r$. The absolute magnitudes in the $NUV$ and $r$ bands were extracted from the NSA catalogue. (4) Star formation rate from \citet{chang15}. (5) Difference between the global kinematic position angles (PA) obtained for H$\alpha$ and stellar velocity fields, $\rm \Delta PA=|PA_{gas}-PA_{stellar}|$. (6) The luminosity of [O\,{\sc iii}]$\lambda5007$ \AA\ measured within the inner 2\farcs5 diameter with MaNGA data. (7) Number of exposures performed for each galaxy. (8) Time of each exposure. (9) Program identification codes on Gemini. (10) -- (11) Spatial resolution for each galaxy measured from the full-width at half maximum (FWHM) of field stars in the GMOS-IFU acquisition image. (12) MaNGA spatial resolution.} 
  \resizebox{\textwidth}{!}{\begin{minipage}{1.19\textwidth}
\small
\begin{tabular}{cccccccccccc}
\toprule
MaNGA-ID & $z$ & $NUV-r$ & log(SFR)  & $\Delta$PA ($^{\circ}$) & L$_{\rm [OIII]}$  & Exposure  & Exposure  & Program ID & GMOS Spatial & GMOS Spatial & MaNGA spatial \\
 & & & (M$_{\odot}$ yr$^{-1}$)& & (10$^{39}$\,erg\,s$^{-1}$)  & Number & Time (s) & &  Resolution (pc) & Resolution ($^{\prime\prime}$) & Resolution (pc) \\
(1) & (2)& (3)&(4) & (5)& (6)  & (7) & (8) & (9) & (10) & (11) & (12)\\
\midrule
1-523238 & 0.0277 & 5.43  & -2.353 & 115.0 &  0.831 & 4 & 1200 & GN-2020A-Q-130 & 400 & 0.70 & 1440\\
1-352045 & 0.0316 & 5.90  & -4.388 &47.0 & 0.370 & 3 & 1200 & GN-2020A-Q-130 & 390 & 0.60 & 1640\\
1-114245 & 0.0288 & 5.94 & -3.758 &  105.0 & 1.068 & 4 & 1000 & GN-2020A-Q-130 & 330 & 0.55 & 1500\\
1-197230 & 0.0752 & 6.02 & -3.448 &  109.0 & 7.304 & 4 & 1200 & GN-2020A-Q-130 & 890 & 0.60 & 3710\\
1-296801 & 0.0531 & 5.73 & -3.528 &  38.5 & 4.619 & 4 & 1200 & GN-2020A-Q-130 & 530 & 0.50 & 2690 \\
1-24104 & 0.0299 & 5.90  & -3.783 &  159.5 & 2.006 & 3 & 1200 & GN-2020A-Q-130 & 380 & 0.61 & 1560\\
1-385124 & 0.0289 & 5.71 & -3.958 &  41.5 &   3.102 & 3 & 970 & GN-2020A-Q-226 & 300 & 0.50 & 1500\\ 
1-474828 & 0.0252 & 5.78 & -2.853 &  38.0 &   0.823 & 3 & 980 & GN-2020A-Q-226 & 300 & 0.57 & 1320 \\ 
1-279073 & 0.0323  & 5.22 & -4.058 &  38.0 & 2.301 & 3 & 980 & GN-2020A-Q-226 & 370 & 0.55 & 1680\\
\bottomrule
\end{tabular} 
 \end{minipage}}
\label{exp}
\end{table*}
\subsection{The GMOS sample}
 The [O\,{\sc iii}]$\lambda5007$ luminosity (computed from the  fluxes measured within a nuclear aperture of 2\farcs5 diameter) and $\rm \Delta PA$ distributions for the Red Geyser sample are shown in Fig. \ref{fig:palum}.
 In order to select the Red Geysers to be observed with GMOS-IFU, we computed the values of $\rm \Delta$PA that divide the Red Geysers of MaNGA sample into three groups, each group containing $\sim$33$\%$ of the total sample. The groups are: $\rm 10^\circ < \Delta PA \leq 40^\circ $ ; $\rm 40^\circ < \Delta PA \leq 105^\circ$ and $\rm 105^\circ < \Delta PA < 170^\circ$. Then, we splitted each group in three [O\,{\sc iii}]$\lambda5007$ luminosity bins:  L$_{\rm [OIII]} <
1.0\times10^{39}$\,ergs\,s$^{-1}$;  L$_{\rm [OIII]} = (1.0-2.45)\times10^{39}$\,ergs\,s$^{-1}$; and  L$_{\rm [OIII]}> 2.45\times10^{39}$\,ergs\,s$^{-1}$. We chose L$_{\rm [OIII]}$ as an indicator of the AGN power and the bins were chosen such that each contain 1/3 of the total sample. The range of luminosities in our sample is L$_{\rm [OIII]}= (0.10-20)\times10^{39}$\,ergs\,s$^{-1}$.  Our final sample contains nine objects, one of each group for each luminosity bin, as shown in the Table \ref{exp} with the MaNGA identification of each galaxy and Fig.\ref{fig:palum}. Thus, covering a wide range of $\rm \Delta PA$ and L$_{\rm [OIII]}$, essential to properly map the gas emission and kinematics of Red Geysers. Table \ref{exp} also presents the properties of the GMOS-IFU sample, such as redshift ($z$), $NUV-r$ and SFR.

Figure \ref{fig:1-385124} shows the optical image, gas velocity field, equivalent width map for one Red Geyser of our sample with MaNGA data. The target clearly presents the criteria for being classified as a Red Geyser. The same properties are shown in Fig.\ref{fig:1-296801}--\ref{fig:1-279073} for other galaxies with extended emission in the GMOS-IFU sample.  We also present in these figures the flux distributions, the excitation maps and the BPT and WHAN diagrams with DAP data, which are discussed in Section \ref{Sec:results} and \ref{discussion}.

\subsection{GMOS observations and data reduction}

The GMOS instruments are installed at the Gemini-North and Gemini-South 8.1 m diameter telescopes. GMOS  allows observations between 4000-10000 \AA\ \citep{smith02}, in the modes of long-slit spectroscopy, multi-object spectroscopy, integral field spectroscopy or imaging \citep{hook04}.
The nine Red Geysers were observed with GMOS integral field unit (IFU) of the Gemini-North telescope. GMOS-IFU can operate in two modes, the \textit{one slit} mode and the \textit{two slit} mode. For our observations, we chose the \textit{one slit} mode, since it produces a broader spectral range when compared to the two slit mode. This allows the inclusion of the strongest optical emission lines, such as H$\beta$, [O\,{\sc iii}]$\lambda5007$, [O\,{\sc i}]$\lambda6300$, H$\alpha$, [N\,{\sc ii}]$\lambda\lambda$6548,6583 and [S\,{\sc ii}]$\lambda\lambda$6716,6730,  which are important to determine the gas excitation mechanism.
In the \textit{one slit} mode, GMOS-IFU uses 750 fibres, each of them  connected to a hexagonal lens, of which 500 are dedicated to the science exposures, while the other 250 fibres cover the sky observations, separated by 1 arcmin from the main object. 
The GMOS-IFU FoV for this mode is 5\farcs0$\times$3\farcs5.
The observations were performed using the B600 grating. With this configuration, GMOS-IFU data have a spectral coverage from $\sim$4400 \AA\ to 7400 \AA.

Table\ref{exp} shows the list of galaxies observed, the number of individual exposures, the exposure time and the Gemini program. For seven galaxies, we have obtained two exposures with the spectra centred at 5900 \AA\ and at least one with the spectra centred at 5850 \AA, in order to interpolate the gaps between the GMOS detectors. For MaNGA 1-296801 and MaNGA 1-197230 the spectra were centred at 6000 \AA, 6005 \AA\ and 6015 \AA, 6020 \AA, to avoid that strong emission lines fall into the detectors gaps. 

The data reduction was performed with {\sc iraf} software using the standard GEMINI package, which includes routines developed specifically for GMOS-IFU data reduction. We have followed the standard steps of spectroscopic data reduction, including bias subtraction, flat-field correction,  background subtraction for each science data, quantum efficiency correction, sky subtraction, wavelength and flux calibration. Furthermore, removal of cosmic rays was performed with the {\sc LACOS} algorithm \citep{van01}. Data cubes for each exposure were created with an angular sampling of 0\farcs05$\times$0\farcs05. 

After the standard reduction processes, we also applied the techniques presented in \citet{menezes19} and \citet{ricci14} in order to improve the quality of the data cubes. First, we corrected all exposures for the differential atmospheric refraction effect using the equations proposed by \citet{Bonsch98} and \citet{Filippenko82}. After this, we created a data cube for each object by calculating the median of all observed exposures. Then, we removed high spatial frequency noises using a Butterworth filter \citep{woods02} with a filter order $n = 2$ and a cut-off frequency of $\sim$ 0.20 $F_{NY}$, where $F_{NY}$ is the Nyquist frequency. These parameters for the filters assures that we are removing only those structures with a spatial frequency that is higher than the frequency of the PSF of the data cubes. Next, we used Principal Component Analysis (PCA) Tomography \citep{steiner09} to remove instrumental fingerprints with spatial and spectral low-frequency signatures. Finally, we removed the telluric lines of all spectra of the data cubes. At this point, the data cubes are ready for science analysis.

\begin{figure}
	\includegraphics[width=0.49\textwidth]{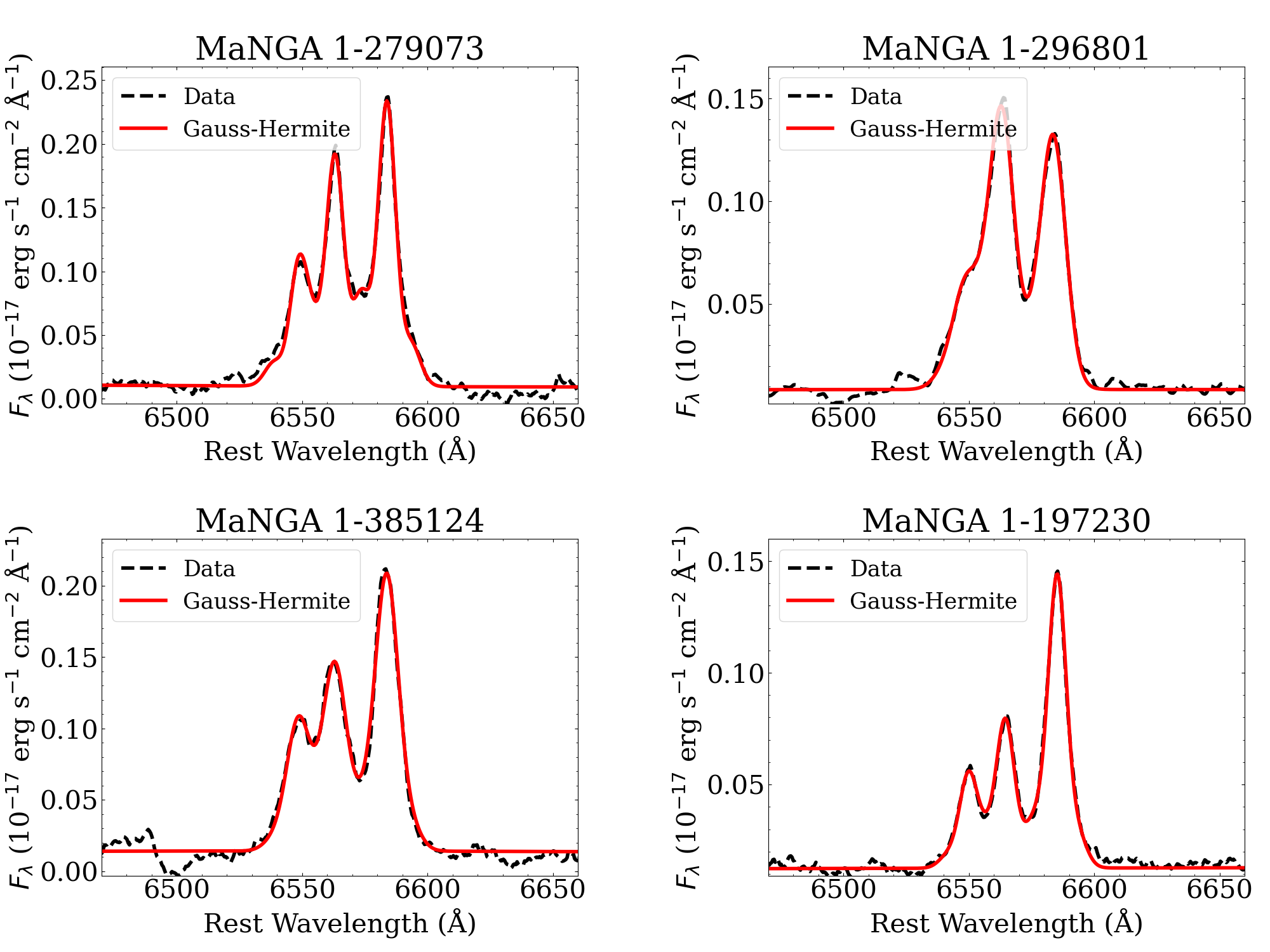}
    \caption{ Examples of fits of the  H$\alpha$ and [N\,{\sc ii}]$\lambda\lambda$6548,6583 emission-line profiles for the nuclear spaxel of four galaxies of our GMOS-IFU sample. The data are shown as black lines and the fits by Gauss–Hermite series as red lines.}
    \label{fig:ifscube}
\end{figure}

The spectral resolution obtained from the full width at half-maximum (FWHM) of the typical emission lines of the CuAr lamp spectrum is $\sim$ 1.6 \AA, which corresponds to $\sim$ 90 km s$^{-1}$. The spatial resolutions shown in Table \ref{exp} were estimated from the measurement of the FWHM of the flux distributions of the field stars and with the distances of galaxies determined by the redshifts of Table \ref{exp}.

\section{Spectral fitting}
\label{sec:spectralfitting}

To subtract the underlying stellar contribution of the GMOS-IFU spectra, we have used the {\sc ppxf} routine to fit the continuum/absorption spectra of each galaxy. {\sc ppxf} assumes that the observed spectrum can be modelled as the convolution between stellar templates and the line-of-sight velocity distribution (LOSVD),  which is represented by a Gauss-Hermite series \citep{cappellari04,cappellari17}. In our fit, four Gauss-Hermite moments were included: velocity, velocity dispersion, $h_{3}$ and $h_{4}$, and {\sc ppxf}  was allowed to use multiplicative Legendre polynomials to correct the continuum shape during the fit. The chosen stellar template library was MILES-HC \citep{westfall19} which includes 42 spectra obtained from the MILES stellar spectra library \citep{falcon11, sanchez06}. The main reason to use this library was its previous application in the DAP data products, which we used to select our sample, thus making our analysis consistent with the MaNGA data.

\begin{figure*}
	\includegraphics[width=0.90\textwidth]{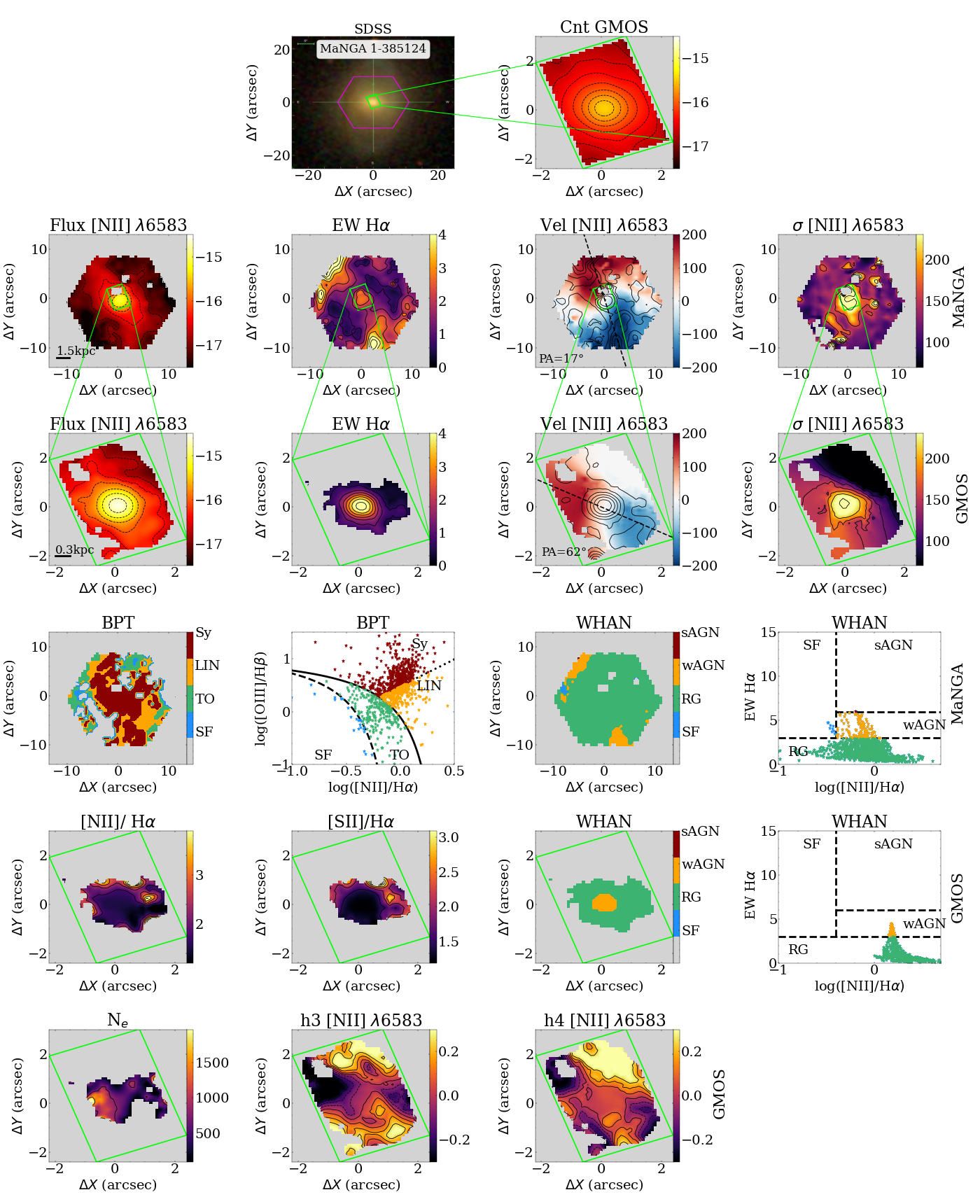}
    \caption{Maps produced using MaNGA-DAP and modelling the emission-line profiles of GMOS-IFU data for the galaxy MaNGA 1-385124. The first row presents the optical image from SDSS with the MaNGA-IFU shown in magenta and the GMOS-IFU FoV shown in green, and the continuum map. The green rectangles in all images indicate the GMOS-IFU FoV. In the second (MaNGA) and third (GMOS-IFU) rows the [N\,{\sc ii}]$\lambda$6583 flux, H$\alpha$ equivalent width (EW), [N\,{\sc ii}]$\lambda$6583  velocity and [N\,{\sc ii}]$\lambda$6583 velocity dispersion maps are shown. The continuum and flux maps are in units of  erg s$^{-1}$ cm$^{-2}$ arcsec$^{-2}$  \AA$^{-1}$ and erg s$^{-1}$ cm$^{-2}$ arcsec$^{-2}$. The black bars in the flux maps show the MaNGA and GMOS-IFU spatial resolutions. The EW maps are in units of \AA. The black contours in the velocity fields represent the [N\,{\sc ii}]$\lambda$6583 equivalent width distribution. The velocity fields are in the unit of km s$^{-1}$ relative to the systemic velocity of the galaxy. In all velocity maps, the dashed black line shows the position angle (PA) of the kinematic major axis  of the gas velocity field. The fourth row shows the spatially resolved BPT excitation map,  BPT diagram \citep{baldwin81}, WHAN map and  WHAN \citep{cid10,cid11} diagram with MaNGA data. The continuous lines shown in the BPT diagrams are from \citet{kewley01}, the dashed line is from \citet{kauffmann03b} and the dotted line is from \citet{cid10}. In the fifth row the [N\,{\sc ii}]/H$\alpha$, [S\,{\sc ii}]/H$\alpha$  emission-line ratios, WHAN excitation map and WHAN diagrams are presented for GMOS-IFU. The following labels were used in the diagrams: LIN: LINER (Low-ionisation nuclear emission-line region), Sy: Seyfert, SF: star-forming region, TO: transition object, wAGN: weak AGN, sAGN: strong AGN and RG: retired galaxy (region). The sixth row shows the electron density map in the units of cm$^{-3}$,  $h_{3}$ and $h_{4}$ Gauss-Hermite moments from GMOS-IFU data. }
   \label{fig:1-385124}
\end{figure*}

After the subtraction of the stellar contribution from the observed data cubes, we used the {\sc ifscube} \citep{ruschel20,ruschel21} Python package to fit the emission-line profiles and measure the gas properties. This package allows the fit of emission-line profiles with Gaussian functions or Gauss-Hermite series. From tentative fits of the emission lines we concluded that their profiles present deviations from a single Gaussian shape, presenting blue or red wings. We thus decided to fit the profiles with Gauss-Hermite series, which can account for the observed deviations. The profiles of the following emission lines were fitted by Gauss-Hermite series: H$\beta$, [O\,{\sc iii}]$\lambda\lambda 4959,5007$, [O\,{\sc i}]$\lambda6300$, H$\alpha$, [N\,{\sc ii}]$\lambda\lambda$6548,6583 and [S\,{\sc ii}]$\lambda\lambda$6716,6730. 

The following constraints were imposed:
%
(i) The kinematics of all emission lines were kept tied. The centroid velocity can range from $-350$ to $350$ km s$^{-1}$ in relation to the velocity calculated using the redshift of each galaxy. The velocity dispersion can vary between $30$ and $350$ km s$^{-1}$;
(ii) Gauss-Hermite moments $h_{3}$ and $h_{4}$ are restricted to the interval -0.3--0.3 for all emission lines. Negative values of $h_{3}$  indicate the presence of a blue asymmetric wing, while positive values indicate the presence of a red asymmetric wing in the line profiles. The $h_4$ moment quantifies symmetric deviations of the line profiles from a Gaussian, with negative (positive) values indicating a flatter (more peaked) line profile -- i.e. profiles with a lower or larger kurtosis compared to a Gaussian function.
(iii) The [O\,{\sc iii}]$\lambda$5007/[O\,{\sc iii}]$\lambda$4959 and [N\,{\sc ii}]$\lambda$6583/[N\,{\sc ii}]$\lambda$6548 flux ratios were fixed to their theoretical values of 2.98 and 3.08 \citep{Osterbrock06}, respectively. These constraints are necessary to properly fit the emission lines in the blue part of the spectra, which are very faint in some objects. 
Figure \ref{fig:ifscube} shows examples of the fits of the [N\,{\sc ii}]$\lambda\lambda$6548,6583 and H$\alpha$ emission-lines in the central spaxel of four galaxies in our sample. One may see that the emission-line profiles are well reproduced by a Gauss-Hermite series.

\section{Results}
\label{Sec:results}
\subsection{Emission-line flux distributions}

Figures \ref{fig:1-385124} and \ref{fig:1-296801}--\ref{fig:fluxt} show the continuum flux distribution from the GMOS-IFU data for all galaxies of our sample, obtained by computing the mean flux values in a $\sim$500\AA\ wide spectral window centred at $\sim$5450\AA. The GMOS-IFU and MaNGA-DAP flux distributions of [N\,{\sc ii}]$\lambda$6583 for the six galaxies with extended emission are shown in Figures \ref{fig:1-385124} and  \ref{fig:1-296801}--\ref{fig:1-279073}. All images were rotated so that the North is up and East is to the left. The light grey  regions in the GMOS-IFU maps correspond to locations where the corresponding emission line is not detected above 3$\sigma$ of the noise level computed in a spectral window next to the line, as well as regions not covered by the GMOS-IFU field of view (delineated by the green lines). The light grey in the DAP maps are regions outside of MaNGA FoV or were removed using the flux quality mask from DAP.

 We do not present the flux maps for the other emission lines because they are similar to those of [N\,{\sc ii}]$\lambda$6583. In some cases we were able to measure  H$\beta$ and [O\,{\sc iii}]$\lambda\lambda$4959,5007 only in a few spaxels. The GMOS signal-to-noise ratio (SNR) in the blue part of the spectra is low (SNR < 10) and  these  lines are weak in Red Geysers, not being detected above 3$\sigma$ the noise level.  Figures \ref{fig:snr}--\ref{fig:snr2} present the evaluated SNR for the blue and red regions of the observed spectra, as well as the extracted spectrum for the nuclear spaxel of each target.

In general for both datasets, [N\,{\sc ii}]$\lambda$6583 flux distributions follow the same pattern in each galaxy, with an emission peak at the nucleus and a weak extended emission over the FoV. The strongest [N\,{\sc ii}]$\lambda$6583 extended emission accross the FoV are observed for the following galaxies: MaNGA 1-385124, 1-296801,1-24104,1-114245, 1-197230 and 1-279073. MaNGA 1-385124, 1-24104 and 1-114245 show extended emission similar to the Akira galaxy \citep{riffel19}. Figure \ref{fig:fluxt} shows the [N\,{\sc ii}]$\lambda$6583 GMOS-IFU flux maps for the three galaxies where the emission lines are detected only in a compact structure with a radius of $\sim$1\farcs0. In these galaxies, the blue emission lines are not detected in the GMOS-IFU spectra throughout the whole FoV.

\subsection{Equivalent width maps}

The H$\alpha$ equivalent width (EW) maps for the six galaxies with extended emission  are presented in the second panels of the second and third row of Figures \ref{fig:1-385124} and \ref{fig:1-296801}--\ref{fig:1-279073}. 
 The DAP EW maps, in general, show the largest values in the nuclear region, and they show an EW bi-symmetric feature aligned with the gradient of the gas velocity field. MaNGA 1-385124, 1-296801 ,1-197230 and 1-279073 show EW H$\alpha$ > 3 \AA\ in some regions.

The GMOS-IFU EW maps clearly show that the bi-symmetric emission features observed in the large scale MaNGA maps extend inward to the nuclear region, except for galaxy MaNGA 1-385124. 
For the galaxy MaNGA 1-197230, this pattern is less evident.  
The GMOS-IFU EW  maps of four galaxies present nuclear region with values larger than 3 \AA\ (MaNGA 1-279073, 1-296801, 1-385124 and 1-197230). In two of these maps, there are locations with EW greater than 6 \AA\ (MaNGA 1-279073 and 1-296801). Only MaNGA 1-24104 and 1-114245 do not present EW larger than 3 \AA\ across the H$\alpha$ EW map, although it presents regions with $\sim$2 \AA. The highest equivalent widths are seen at the nucleus, with the intermediate/high values distributed along with the bi-symmetric pattern. GMOS-IFU EW maps values are usually larger than those of the MaNGA-DAP map. The [N\,{\sc ii}] EW distributions are not shown, because they present a similar pattern. 

\citet{riffel19} found a change in the orientation of the ionisation  pattern from the nuclear region to galactic scales for the Akira galaxy, the prototype Red Geyser. Akira also presents a variation in the outflow orientation observed in the gas velocity fields. Both variations are interpreted by \citet{riffel19} as due to precession of the accretion disc. Among the six Red Geysers with extended emission over the whole GMOS-IFU FoV (Fig. \ref{fig:1-385124} and Fig. \ref{fig:1-296801}--\ref{fig:1-279073}), at least three galaxies show misalignment of the gas ionisation pattern from the nucleus to kpc scales (MaNGA 1-279073, 1-114245, 1-24104), while the other three (MaNGA 1-385124 and 1-197230, 1-296801) present similar emission orientations at nuclear and kpc scales. In Sections \ref{kinematics1} and \ref{kinematics2}, we quantify and discuss the misalignment of velocity fields and the implications of the result.

\subsection{GMOS emission-line ratios}

Figures \ref{fig:1-385124} and \ref{fig:1-296801}--\ref{fig:1-279073} also show the [N\,{\sc ii}]$\lambda$6583/H$\alpha$ and [S\,{\sc ii}]$\lambda \lambda$6716,6730/H$\alpha$ emission-line ratio maps for the six galaxies with GMOS-IFU extended emission. 
These maps were used to verify the gas excitation mechanisms. The emission-line ratios show a wide range of values and distributions for each galaxy. The  [N\,{\sc ii}]$\lambda$6583/H$\alpha$ values range from 0.75 (for MaNGA 1-279073) to 4.5 (MaNGA 1-197230), while [S\,{\sc ii}]$\lambda\lambda$6717,6731/H$\alpha$ shows values ranging from 0.75 to 3. 

For three objects the  [N\,{\sc ii}]$\lambda$6583/H$\alpha$ ratio has the lowest values in the nucleus: $\sim$1 for MaNGA 1-296801,  $\sim$1.5  for  1-385124 and $\sim$2 for 1-197230. The highest values are in ``shells'' around the nucleus (greater than 2 for MaNGA 1-296801, $\sim$3 for MaNGA 1-385124 and greater than 3.5 for  1-197230) observed at the edges of the distributions. The targets MaNGA 1-24104, 1-114245 and 1-279073 show [N\,{\sc ii}]$\lambda$6583/H$\alpha$ emission-line ratios with the largest values at the nucleus, but also present extra-nuclear knots of high values. While for [S\,{\sc ii}]$\lambda \lambda$6716,6730/H$\alpha$,  smaller values in the nuclear region are observed, surrounded by higher values, for  MaNGA 1-385124, 1-296801, 1-279073. For MaNGA 1-24104 and 1-114245 the largest  [S\,{\sc ii}]/H$\alpha$ values are observed along the EW bi-symmetric feature. Unlike [N\,{\sc ii}]/H$\alpha$, the distribution of [S\,{\sc ii}]/H$\alpha$ show higher values at the nucleus for MaNGA 1-197230. The [O\,{\sc i}]$\lambda$6300/H$\alpha$ maps are not shown but are used to construct the BPT diagram in Fig. \ref{fig:bpt}.

\subsection{GMOS electron density maps}

Assuming a temperature of 10$^{4}$ K, the [S\,{\sc ii}]$\lambda $6716/[S\,{\sc ii}]$\lambda$6730 ratio was used to  obtain the electron density ($N_{e}$) using the {\sc PyNeb} python routine \citep{luridiana15}.  
The $N_{e}$ maps are show in Figures \ref{fig:1-385124} and \ref{fig:1-296801}--\ref{fig:1-279073} and present values in the range 100 cm$^{-3}$ $<$ $N_{e}$ $<$ 2000 cm$^{-3}$. The highest $N_{e}$ values are usually observed at the edges of the FoV. The only exception is MaNGA 1-279073 that shows the greatest values ($N_{e} >$ 1000 cm$^{-3}$) at the nucleus and at the edges of the FoV. 
 
 The galaxy MaNGA 1-296801 presents electron density greater than 3000 cm$^{-3}$ to the south-west and north-east of the nucleus, being the object with the highest $N_{e}$ in our sample. MaNGA 1-385124 presents $N_{e} >$ 1200 cm$^{-3}$ to the east of the nucleus, and  1-114225 shows $N_{e} >$ 600 cm$^{-3}$ to the north. For  1-19720, it was possible to determine $N_{e}$ only in a few spaxels. Finally, MaNGA 1-24104, on the other hand, shows some ``hotspots'' where the highest densities ($N_{e} >$ 600 cm$^{-3}$) are observed.

\subsection{Gas kinematics}
\label{kinematics1}
We  present the [N\,{\sc ii}]$\lambda$6583 velocity and velocity dispersion ($\sigma$) maps from GMOS-IFU and DAP data for the six galaxies with extended emission, as shown in the second and third rows of Fig. \ref{fig:1-385124} and Fig. \ref{fig:1-296801}--\ref{fig:1-279073} . The $h_{3}$ and $h_{4}$ maps for GMOS-IFU are shown in the sixth row of Fig. \ref{fig:1-385124} and Figs. \ref{fig:1-296801}--\ref{fig:1-279073}. Similarly to GMOS-IFU flux and EW maps, the light grey regions are locations where the emission line was not detected above 3$\sigma$ the noise level. The light grey regions in the DAP maps are regions outside of MaNGA FoV or removed using the data quality mask from DAP. The black contours in the velocity fields represent the [N\,{\sc ii}]$\lambda$6583 equivalent width values distribution. The systemic velocity of the galaxies was subtracted for all velocity maps.

The MaNGA velocity fields reach absolute values greater than  $ 200-300$ km s$^{-1}$ and GMOS-IFU maps reach absolute values larger than $150-300$ km s$^{-1}$. The velocity fields at small and large scales are distinct with the orientations changing.  We have used the {\sc fit$_-$kinematic\_pa} routine of \cite{krajnovic06} to symmetrize both gas velocity fields and to measure the orientation of the line of nodes. The corresponding PA values for each galaxy are plotted in Fig. \ref{fig:1-385124} and Figs. \ref{fig:1-296801}--\ref{fig:1-279073}, clearly showing that the gas velocity fields have different orientations at small and large scales. The orientation change is between 12$^{\circ}$ and 60$^{\circ}$ from the nuclear region (GMOS-IFU) to kpc scales (MaNGA), as shown in Table \ref{pagmosmanga}.

\begin{table}
\centering
\caption{Large and small scale kinematic position angle values. (1) Galaxy identification in the MaNGA survey. (2) GMOS kinematic position angle. (3)   Difference between the GMOS and MaNGA global kinematic position angles (PA).  }
\small
\begin{tabular}{cccc}
\toprule
MaNGA-ID & PA  & PA  & $\Delta$PA \\
 & MaNGA & GMOS & |MaNGA-GMOS| \\
\midrule
1-114245 & 6.0$^{\circ}\pm 0.5^{\circ}$ & 35.0$^{\circ}\pm 0.5^{\circ}$ & 29.0$^{\circ}\pm0.7^{\circ}$ \\
1-197230 & 17.0$^{\circ}\pm 0.5^{\circ}$  & 159.0$^{\circ}\pm 2.0^{\circ}$ & 38.0$^{\circ}\pm2.0^{\circ}$ $^{\dagger}$  \\
1-296801 & 128.0$^{\circ}\pm 0.5^{\circ}$  & 140.0$^{\circ}\pm 1.0^{\circ}$ & 12.0$^{\circ}\pm1.0^{\circ}$\\
1-24104 & 24.0$^{\circ}\pm 1.0^{\circ}$  & 144.0$^{\circ}\pm 1.0^{\circ}$ & 60.0$^{\circ}\pm1.5^{\circ}$ $^{\dagger}$\\
1-385124 & 17.0$^{\circ}\pm 0.5^{\circ}$  & 62.0$^{\circ}\pm 3.5^{\circ}$ & 45.0$^{\circ}\pm3.5^{\circ}$\\
1-279073 & 155.0$^{\circ}\pm 0.5^{\circ}$  & 122.0$^{\circ}\pm 5.0^{\circ}$ & 33.0$^{\circ}\pm5.0^{\circ}$ \\
\bottomrule
\end{tabular} 

 $\dagger$ Note that PA is the angle along which the velocity shows the maximum gradient, then $\rm \Delta PA$ needs to be corrected for 180-$\rm \Delta PA$ for these two galaxies.
\label{pagmosmanga}
\end{table}

The MaNGA velocity dispersion maps reveal $\sigma > 200-250$ km s$^{-1}$ at the nucleus and for the regions along the bi-symmetric emission patterns, indicating disturbance in the gas velocity fields. Regions of intermediate  $\sigma$  values 150$-$200 km s$^{-1}$ surround the regions with the largest values. The lowest $\sigma$ [N\,{\sc ii}]$\lambda$6583 are between 70$-$150 km s$^{-1}$ and are found in ``spots'' across the FoV. The GMOS data show the highest $\sigma$  values observed co-spatially with the nucleus for all galaxies, reaching up to $ \sigma \sim 200-250$ km s$^{-1}$. The $\sigma$ values $\sim$150$-$200 km s$^{-1}$ are commonly seen in narrow strips across the FoV, except for the galaxies 1-197230 and 1-279073. The first galaxy shows $\sigma \sim$200 km s$^{-1}$ almost in the entire velocity dispersion field. For 1-279073, the intermediate $\sigma$ values are observed near to the nucleus.

The $h_{3}$ Gauss-Hermite moment reproduces asymmetric profiles, $h_{3}$ positive indicating a red asymmetric wing in the emission-line, and $h_{3}$ negative indicating a blue asymmetric wing \citep{riffel10}. The $h_{4}$ moment measures whether the profile has a flatter kurtosis ($h_{4}<0$) than a Gaussian profile, or a more peaked kurtosis with $h_{4}>0$ \citep{riffel10,ruschel21}. All maps reveal $h_{4}$ usually positive in almost the entire FoV with different distributions for each galaxy. Usually, the positive $h_{3}$ are observed co-spatially with the blueshifted positions in the velocity fields and the negative values with the redshifted positions. Only MaNGA 1-296801 shows $h_{3}>0$ in both regions, but the $h_{3}$  map seems dominated by negative $h_{3}$.

\section{DISCUSSION}
\label{discussion}
\subsection{Gas excitation}

Optical emission-line ratio diagrams, such as the BPT  diagrams \citep{baldwin81}, can be used to map the excitation of the gas. In the case of the [O\,{\sc iii}]$\lambda$5007/H$\beta$ $\times$ [N\,{\sc ii}]$\lambda$6583/H$\alpha$ diagram, the regions are divided among SF  corresponding to gas ionised by young stars, Transition Objects (TO) where ionisation is due to AGN together with star formation, and LINERs/Seyfert. These divisions follow the criteria of \citet{kauffmann03b} and \citet{kewley01,kewley06}, where regions with

\begin{equation} 
\label{eq1}
\rm log([O\:III]\lambda5007/H\beta) \: < \: \frac{0.61}{log([N\:II]\lambda6583/H\alpha)-0.05}+1.3
\end{equation}
\noindent correspond to SF galaxies, while those defined by
\begin{equation}
\label{eq2}
\rm log([O\:III]\lambda5007/H\beta) \: > \: \frac{0.61}{log([N\:II]\lambda6583/H\alpha)-0.47}+1.19
\end{equation}

\noindent are LINERs or Seyferts. Objects between these two curves are classified as TO. Following \citet{cid10}, we separate LINER and Seyfert excitation according to the division line

\begin{equation}
\label{eq3}
\rm log([O\:III]\lambda5007/H\beta) \: = \: 1.01log([N\:II]\lambda6583/H\alpha)+0.48
\end{equation}
\noindent where Seyfert (LINER) objects are located above (below) this line.

\begin{table*}
\centering
\caption{ Emission-line ratios and H$\alpha$ equivalent width values extracted within a nuclear aperture equal to GMOS-IFU spatial resolution -- column (11) of Table \ref{exp}-- and 2\farcs5 for MaNGA. (1) Galaxy identification in the MaNGA survey. (2)--(8) The emission-line ratios using MaNGA data.  (3)--(9) Same as (2)--(8), but with GMOS-IFU data. (10) The H$\alpha$ equivalent width with MaNGA data. (11) Same as column (10), but with GMOS-IFU data. (12) Electron density values using the integrated spectra within a nuclear aperture equal to GMOS-IFU spatial resolution. (13)  Electron density lower limit using the integrated spectra within a nuclear aperture equal to GMOS-IFU spatial resolution. (14) Electron density upper limit using the integrated spectra within a nuclear aperture equal to GMOS-IFU spatial resolution. (15) AGN or non-AGN using  H$\alpha$ EW $>$ 1.5\AA\ for MaNGA data. (16) AGN or non-AGN using  H$\alpha$ EW $>$ 1.5\AA\ for GMOS-IFU data.}
\resizebox{\textwidth}{!}{\begin{minipage}{1.37\textwidth}
\small
\begin{tabular}{ccccccccccccccccc}
\toprule
MaNGA-ID & [N\,{\sc ii}]/H$\alpha$ & [N\,{\sc ii}]/H$\alpha$ & [O\,{\sc iii}]/H$\beta$ & [O\,{\sc iii}]/H$\beta$ &  [O\,{\sc i}]/H$\alpha$ & [O\,{\sc i}]/H$\alpha$ & [S\,{\sc ii}]/H$\alpha$ & [S\,{\sc ii}]/H$\alpha$ & H$\alpha$ EW & H$\alpha$ EW  & $N_{e}$ & $N_{e_down}$ & $N_{e_up}$ & AGN & AGN\\
 & MaNGA & GMOS & MaNGA & GMOS & MaNGA & GMOS & MaNGA & GMOS & MaNGA & GMOS & GMOS & GMOS & GMOS & MaNGA & GMOS\\
(1) & (2) & (3) & (4) & (5) & (6) & (7) & (8) & (9) & (10) & (11) & (12) & (13) & (14) & (15) & (16)\\
\midrule
1-279073 & 1.18$\pm$0.01 & 1.27$\pm$0.07 & 2.02$\pm$0.03 & 1.45$\pm$0.39 & 0.34$\pm$0.01 & 0.46$\pm$0.05 & 1.1$\pm$0.01 & 1.13$\pm$0.12 & 3.2$\pm$0.12 & 6.57$\pm$0.27 & 947 & 461 & 1978 & yes & yes\\
1-296801 & 0.79$\pm$0.01 & 0.9$\pm$0.04 & 1.74$\pm$0.03 & 1.08$\pm$0.28 & 0.26$\pm$0.01 & 0.35$\pm$0.03 & 1.05$\pm$0.01 & 1.26$\pm$0.06 & 5.4$\pm$0.09 & 12.01$\pm$0.41 & 321 & 223 & 437 & yes & yes\\
1-385124 & 1.26$\pm$0.01 & 1.53$\pm$0.09 & 5.7$\pm$0.13 & 9.91$\pm$8.37 & 0.37$\pm$0.01 & 0.54$\pm$0.05 & 0.99$\pm$0.01 & 1.21$\pm$0.1 & 2.5$\pm$0.14 & 4.19$\pm$0.15 & 752 & 422 & 1304 & yes & yes\\
1-114245 & 2.14$\pm$0.01 & 1.93$\pm$0.21 & 14.28$\pm$1.44 & -- & -- & 0.18$\pm$0.05 & 1.45$\pm$0.01 & 1.45$\pm$0.19 & 1.19$\pm$0.2 & 1.87$\pm$0.19 & 339 & 95 & 750 & no & yes\\
1-197230 & 1.94$\pm$0.01 & 1.96$\pm$0.09 & 2.94$\pm$0.07 & 2.47$\pm$1.38 & 0.24$\pm$0.01 & 0.25$\pm$0.04 & 1.3$\pm$0.01 & 1.36$\pm$0.09 & 2.62$\pm$0.14 & 4.32$\pm$0.17 & -- & -- & -- & yes & yes\\
1-24104 & 1.34$\pm$0.01 & 1.65$\pm$0.17 & 6.22$\pm$0.29 & 1.57$\pm$0.91 & 0.02$\pm$0.01 & 0.09$\pm$0.04 & 1.2$\pm$0.01 & 1.56$\pm$0.17 & 1.56$\pm$0.17 & 1.83$\pm$0.16 & 324 & 108 & 655 & yes & yes\\
1-474828 & 1.19$\pm$0.03 & 1.39$\pm$0.31 & 46.29$\pm$19.57 & -- & 0.25$\pm$0.02 & -- & 0.93$\pm$0.03 & 0.97$\pm$0.37 & 0.56$\pm$0.31 & 0.48$\pm$0.08 & -- & -- & -- & no & no\\
1-352045 & 1.02$\pm$0.03 & 1.32$\pm$0.73 & 13.77$\pm$6.14 & -- & 0.06$\pm$0.02 & -- & 0.58$\pm$0.03 & 0.8$\pm$0.63 & 0.51$\pm$0.31 & 0.37$\pm$0.15 & -- & -- & -- & no & no\\
1-523238 & 1.72$\pm$0.04 & 1.28$\pm$0.54 & 4.26$\pm$0.19 & 0.86$\pm$0.74 & 0.13$\pm$0.03 & -- & 1.16$\pm$0.04 & 1.17$\pm$0.51 & 0.58$\pm$0.3 & 0.58$\pm$0.22 & -- & -- & -- & no & no\\
\bottomrule
\end{tabular} 
\label{results}
 \end{minipage}}
\end{table*}

With MaNGA data, we construct the BPT diagram and excitation maps using the [N\,{\sc ii}]$\lambda$6583/H$\alpha$ and [O\,{\sc iii}]$\lambda$5007/H$\beta$ emission-line ratios, as shown in fourth row of Figures \ref{fig:1-385124} and \ref{fig:1-296801}--\ref{fig:1-279073}. All galaxies show AGN ionised gas signatures in the BPT excitation maps with MaNGA data. Some of them also have gas ionisation due to star formation, such as MaNGA 1-24425 and 1-296801. The nuclear region (2\farcs5 in diameter) shows LLAGN in all objects of our sample. In the BPT diagrams, gas ionisation due to hot low-mass evolved stars \citep{stasinska08,cid11}, excitation by shocks and true AGN can all result in spectral line signatures that locate the galaxy spectrum in the LINER region. The combination of the  [N\,{\sc ii}]$\lambda$6583/H$\alpha$ emission-line ratio  and the H$\alpha$ equivalent width, known as WHAN diagram, can be used to separate true AGN from ``fake'' AGN  \citep{cid10,cid11}. 

Using the values obtained for H$\alpha$ equivalent width and [N\,{\sc ii}]$\lambda$6583/H$\alpha$ emission line ratio, we built the WHAN diagram \citep{cid10,cid11} and the excitation map using the MaNGA data. They allow identifying if the gas excitation is caused by star formation (H$\alpha$ EW $>$ 3 \AA\ and $\rm log($[N\,{\sc ii}]/H $\alpha) < -0.4$), or by low-mass hot evolved stars  (H$\alpha$ EW $<$ 3 \AA), typical of retired galaxies (RG, \citealt{stasinska08,cid11}). Furthermore, it is also possible to check if the ionisation source is a strong AGN (i.e. Seyfert; H$\alpha$ EW $>$ 6 \AA\ and  $\rm log($[N\,{\sc ii}]/H $\alpha) > -0.4$) or a weak AGN (i.e. LINER; H$\alpha$ EW $>$ 3 \AA\ and $\rm log($[N\,{\sc ii}]/H$ \alpha ) > -0.4$). The fourth row of  Figure \ref{fig:1-385124} and Figs. \ref{fig:1-296801}--\ref{fig:1-279073} show WHAN maps and the respective diagrams using DAP data for the six Red Geysers in our sample that present extended emission. In the nuclear region, within 2\farcs5 in diameter, the gas excitation is due to AGN only in the galaxies MaNGA 1-279073 and 1-296801.

 We  present the maps with the main emission-line ratios using the GMOS-IFU data in the bottom rows of Fig. \ref{fig:1-385124} and Figs. \ref{fig:1-296801}--\ref{fig:1-279073}. Unfortunately, we were not able to detect extended emission in the  [O\,{\sc iii}]$\lambda$5007 and H$\beta$ lines in most objects due to the lower quality of the GMOS-IFU data in the blue part of the spectra. A value of [N\,{\sc ii}]/H$\alpha$ $>$ 1.0 is consistent with gas excited by a central AGN \citep{kewley01,kewley06}. Five galaxies have  [N\,{\sc ii}]/H$\alpha$ greater than 1.0\sout; only the galaxy MaNGA 1-296801 has smaller values ($\sim$1.0). For [S\,{\sc ii}]/H$\alpha$ $>$0.7 the gas excitation may be dominated by AGN photoionisation \citep{kewley01,kewley06}. All galaxies with extended emission show [S\,{\sc ii}]/H$\alpha$ greater than 0.7. Thus, the observed line-ratios are compatible with the presence of low-luminosity AGNs in all galaxies of our sample of Red Geysers.

\begin{figure*}
	\includegraphics[width=0.49\textwidth]{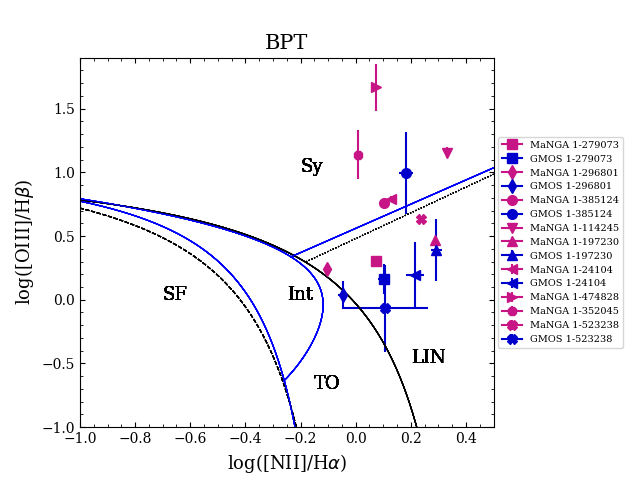}
    \includegraphics[width=0.49\textwidth]{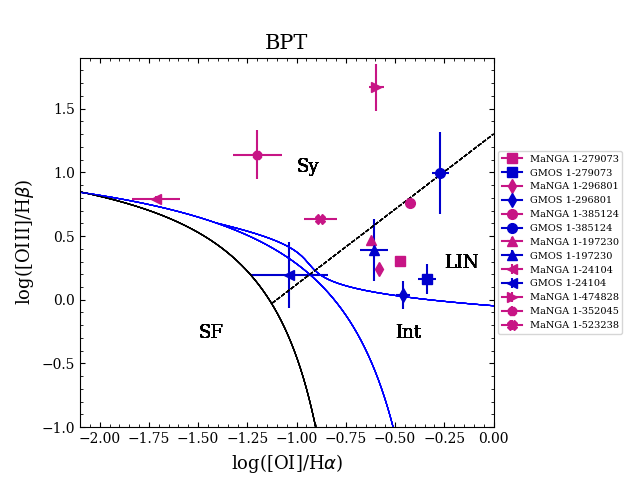}
   \includegraphics[width=0.49\textwidth]{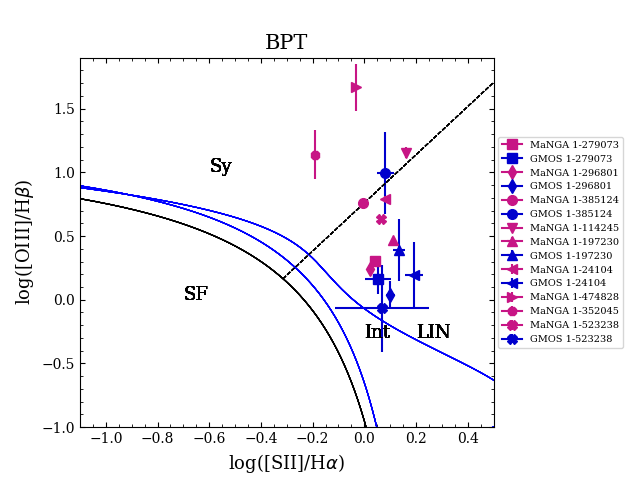}
    \includegraphics[width=0.49\textwidth]{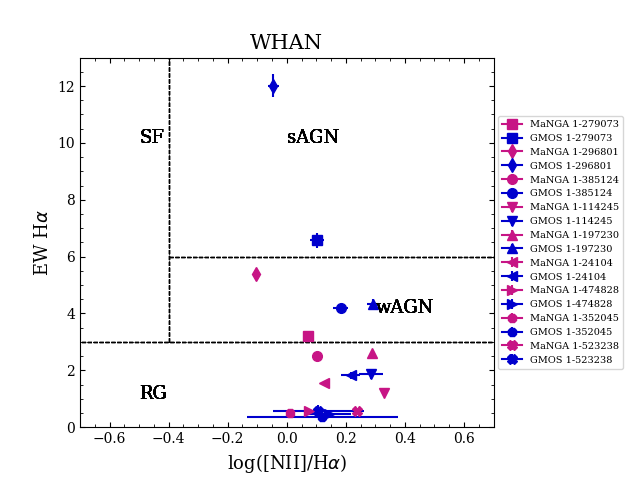}
    \caption{BPT \citep{baldwin81} and WHAN diagrams  plotted with the emission-line ratios within an aperture of 2\farcs5 in diameter for the MaNGA data and within an aperture equal to GMOS-IFU spatial resolution for the corresponding data. The values obtained from MaNGA are shown in magenta and the values from GMOS in blue. The continuous lines shown in the BPT diagrams are from \citet{kewley01}. For the BPT diagram including [N\,{\sc ii}]/H$\alpha$: The dashed line is from \citet{kauffmann03b} and the dotted line is from \citet{cid10}. For the BPT diagrams including  [O\,{\sc i}]/H$\alpha$ and [S\,{\sc ii}]/H$\alpha$: The dashed lines are from \citet{kewley06}. The solid blue lines are from \citet{law21}. The following labels were used in the diagrams: LIN--LINER (Low-ionisation nuclear emission-line region); Sy--Seyfert; SF-- star-forming galaxies; TO--transition objects; wAGN--weak AGN; sAGN--strong AGN; RG--retired galaxies; and Int--intermediate, as defined by \citet{law21}.}
    \label{fig:bpt}
\end{figure*}

Although we do not have measurements of [O\,{\sc iii}]$\lambda$5007 and H$\beta$ for the entire GMOS-IFU FoV, we have integrated the spectra within a nuclear aperture corresponding to the GMOS-IFU spatial resolution in order to increase the signal to noise ratio of the spectra and measure the [O\,{\sc iii}]$\lambda$5007/H$\beta$ nuclear line ratio. This aperture corresponds to 0.3--0.9 kpc at the distance of the galaxies.
We fit the emission-line profiles of the resulting spectrum following the procedure described in Sec.~\ref{sec:spectralfitting}. The resulting fluxes are used to construct the BPT diagrams shown in Fig. \ref{fig:bpt}. This figure also shows the ratios  based on MaNGA data for an aperture of 2\farcs5 diameter (the angular resolution of the data), corresponding to 1.3--3.7 pc at the distance of galaxies.
From the GMOS data, we were able to measure the [O\,{\sc iii}]/H$\beta$ and [O\,{\sc i}]/H$\alpha$  ratios in six galaxies,  while the [N\,{\sc ii}]/H$\alpha$ and [S\,{\sc ii}]/H$\alpha$ were obtained for all galaxies. Columns (2)--(9) of Table \ref{results} show the ratios between the emission lines within the nuclear aperture corresponding to the GMOS-IFU spatial resolution and for an aperture of 2\farcs5 diameter for the MaNGA data.

The [O\,{\sc i}]/H$\alpha$, [N\,{\sc ii}]/H$\alpha$ and   [S\,{\sc ii}]/H$\alpha$ emission-line ratios for the GMOS-IFU data are usually equal to or larger than those obtained for MaNGA, likely because in these objects the peak of the line ratios are observed at the nucleus and mixed with lower values from extranuclear regions in the MaNGA data. On the other hand, the [O\,{\sc iii}]/H$\beta$ values are smaller for GMOS-IFU. This is likely due to the lower signal-to-noise ratio (SNR < 10) in the blue region of the GMOS spectra, as presented in Fig.\ref{fig:snr}--\ref{fig:snr2} and a possible second order contamination of the spectra which is more important in the blue part of the GMOS spectra. From the GMOS-IFU data, we found that all objects are located in the Seyfert and LINER regions of the BPT diagram, except MaNGA 1-296801 which lies in the division between TO and LINER. These results 
were also observed for the MaNGA data. In the BPT diagrams of Fig.\ref{fig:bpt}, we also plot the classification lines from \citet{law21}, the solid blue lines, which uses MaNGA data to improve the boundaries used to determined the gas ionisation mechanism in galaxies. This classification divides the BPT diagrams in four regions: AGN (that we refer to as ``Seyfert''), ; LI(N)ER; star-forming and intermediate (Int). Based on this division, galaxies have Seyfert or LI(N)ER ionisation with both data, except 1-24104 in the [O\,{\sc i}]/H$\alpha$ diagram.

In the WHAN maps and in the respective diagrams of Fig. \ref{fig:1-385124} and Fig. \ref{fig:1-296801}--\ref{fig:1-279073}, all galaxies show regions typical of ``retired galaxies" (H$\alpha$ EW $<$ 3 \AA) and for two of them this is the case in all spaxels. Four Red Geysers have gas ionisation caused by AGN: MaNGA 1-279073, 1-296801, 1-385124 and 1-197230. We also plot this diagram with H$\alpha$ equivalent width values and the [N\,{\sc ii}]/H$\alpha$ ratios obtained within an aperture equal to GMOS-IFU spatial resolution and 2\farcs5 (MaNGA), as shown in Figure \ref{fig:bpt}. The H$\alpha$  EW values are in the columns (10)--(11) of the Table  \ref{results}. \citet{cid10,cid11} proposed a cutoff above 3 \AA\ in H$\alpha$ EW  to identify a galaxy as AGN. According to Fig. \ref{fig:bpt} this is found in two galaxies in the MaNGA data and in four galaxies in the GMOS-IFU data.

 \citet{sanchez18}  have proposed a lower limit of H$\alpha$ EW = 1.5 \AA\ to identify AGN in the MaNGA survey. These authors followed \citet{cid10,cid11}, but relaxing the EW H$\alpha$  Seyfert/LINER borderline to include weaker AGN. Indeed, as discussed in \citet{cid11}, accreting black holes can still contribute with a significant fraction of the the ionizing power for H$\alpha$ EW  between 1.0 and 3.0 \AA. In addition, radio observations of Red Geysers show that they present higher luminosities than control galaxies, indicating that they host LLAGNs \citep{roy21c}. Thus, we perform a second classification of the ionisation source in our sample of Red Geysers adopting  the same H$\alpha$ EW cut used by \citet{sanchez18} to allow the detection of weaker AGN, which are expected to be present as indicated by the radio observations mentioned above.  
Using this lower H$\alpha$ EW value, the number of galaxies with gas ionisation due to AGN rises from two to five in the MaNGA data, and from four to six in the GMOS-IFU data. The following Red Geysers have an AGN identified with both data: MaNGA 1-279073, 1-296801, 1-385124, 1-197230 and 1-24104. Meanwhile,  galaxy 1-114245 is classified as an optical AGN host only in GMOS-IFU data. These results are summarised in columns (15) and (16) of Table \ref{results}.

\begin{figure}
\centering
	\includegraphics[width=0.49\textwidth]{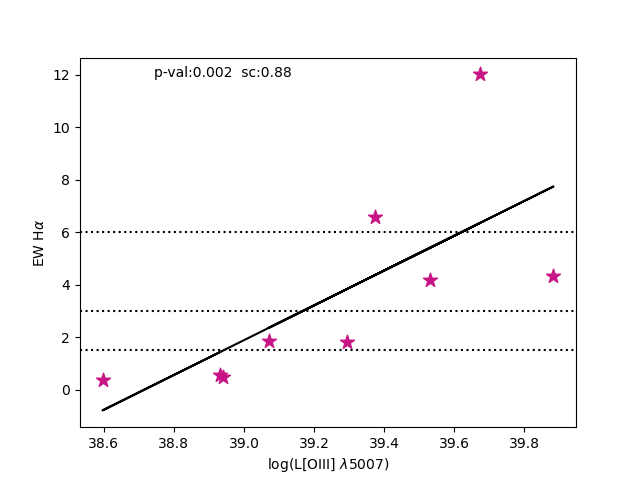}
    \caption{ [O\,{\sc iii}]$\lambda5007$ luminosity (L$_{\rm [OIII]}$) versus H$\alpha$ EW for the nine galaxies observed with GMOS. The luminosity values were extracted from MaNGA data for an aperture of 2\farcs5 of diameter, while the H$\alpha$ EW were determined using GMOS-IFU data with an aperture equal to the angular resolution of Table \ref{exp}.}
    \label{fig:ewlum}
\end{figure}

Using the GMOS emission-line ratios and EW greater than 3 \AA\, we found that at least four Red Geysers host a LLAGN in the nuclear region. Using the criterion  EW$>$ 1.5 \AA\ instead of 3 \AA\ results in an increase of this number  to six objects. The fact that the H$\alpha$ equivalent widths are low for a significant number of Red Geysers is suggestive of the presence of very low luminosity AGN. To check this hypothesis, we compare the H$\alpha$ equivalent widths with the [O\,{\sc iii}]$\lambda$5007 luminosity, a proxy for the bolometric luminosity of the AGN \citep{Heckman04}. This comparison is presented in Fig. \ref{fig:ewlum}. The correlation between these parameters is indeed very significant with a Spearman correlation coefficient of 0.88 and a p-value of 0.002. This test reinforces the view that the low H$\alpha$ equivalent widths that we find for our sample of Red Geysers is a result of the low luminosity AGN that they contain. \citet{riffel19} analysed the gas ionisation source in the Akira galaxy using GMOS-IFU data and also concluded that there is an AGN at the nucleus. Our results also support the scenario that these galaxies host a LLAGN as suggested by \citet{cheung16}. 

\citet{roy18} identified an excess of radio emission in a sample of Red Geysers when compared to non-active galaxies, which was interpreted as due to low-luminosity radio-mode AGNs.  They found that Red Geysers present a mean radio luminosity at 1.4 GHz of L$_{1.4GHz} \sim$2.0$\times$10$^{21}$ W Hz$^{-1}$. As discussed by these authors, for such luminosity be due to star formation, a SFR $\sim$1 M$_{\odot}$ yr$^{-1}$  would be required, but the Red Geysers present much lower SFR, of only 0.001--0.01 M$_{\odot}$ yr$^{-1}$. Therefore, star formation cannot explain their central radio emission, supporting the presence of low-luminosity AGN in Red Geysers. Among the nine galaxies in our sample, three objects are detected in the Very Large Array (VLA) Faint Images of the Radio
Sky at Twenty Centimeters (FIRST) survey \citep{becker95}: MaNGA 1-24104 (L$_{1.4GHz} \sim$0.35$\times$10$^{22}$ W Hz$^{-1}$), MaNGA 1-279073 (L$_{1.4GHz} \sim$0.52$\times$10$^{22}$ W Hz$^{-1}$) and  1-296801 (L$_{1.4GHz} \sim$49.7$\times$10$^{22}$ W Hz$^{-1}$), providing additional support that they host AGNs.

\citet{rembold17} selected an AGN sample from the MaNGA MPL-5 using the diagnostic diagrams described above, which was updated with data from the MPL-8 (the same used here) in \citet{rogerio21} and \citet{deconto21}. This selection is based on the nuclear spectra of the SDSS-III using emission-line fluxes from \citet{tomas13}. Comparing our GMOS Red Geyser sample with \citet{deconto21}, we find that only the galaxy MaNGA 1-279073 is in both samples.  As Red Geysers seem to host LLAGN, they are not detected using the 3\farcs0 diameter fibre from SDSS-III data, and thus our results indicate that the identification of the AGNs in Red Geysers using optical data may require better angular resolution than that provided by SDSS-III and MaNGA data. 

\subsection{ Electron density}
The electron density ($N_{e}$) maps show a wide range  of values: 100--3000 cm$^{-3}$ (Fig. \ref{fig:1-385124} and Fig. \ref{fig:1-296801}--\ref{fig:1-279073}). This result is in agreement with \citet{kakkad18}, \citet{freitas18} and \citet{ruschel21} that using IFS for AGN samples found $N_{e}$ between 100--2500 cm$^{-3}$. We derived the electron density with [S\,{\sc ii}] emission lines integrated within a nuclear aperture equal to GMOS-IFU spatial resolution, as shown in column (12) of the Table \ref{results}. The columns (13) and (14)  are the lower and upper limits for the electron density estimated using the uncertainties in the [S\,{\sc ii}] ratios. The $N_{e}$  range is between 321 cm$^{-3}$ and 947 cm$^{-3}$ in column (12) of the Table \ref{results}. \citet{ruschel21} obtained $\sim$ 800 cm$^{-3}$ for the outflow component and 300 cm$^{-3}$ for regions without outflow. \citet{davies20} determined an average $N_{e}$ of 350 cm$^{-3}$ for active galaxies and 190 cm$^{-3}$ for inactive galaxies with [S\,{\sc ii}] emission lines. All AGN host galaxies in our sample have electron densities greater than 300 cm$^{-3}$  within a nuclear aperture equal to GMOS-IFU spatial resolution. \citet{cheung16} estimated a $N_{e}$ of 100 cm$^{-3}$ for the Akira galaxy, indicating low-density outflows in Red Geysers, but our results show higher $N_{e}$. Furthermore, there are several values greater than 700 cm$^{-3}$  outside the aperture (Fig. \ref{fig:1-385124} and Fig. \ref{fig:1-296801}--\ref{fig:1-279073}) that may be associated with the presence of high-density outflows in these galaxies. Thus, the $N_e$ values in our sample are consistent with previous measurements of the density in ionised outflows, providing additional support that Red Geysers present large scale AGN winds.

\subsection{Gas kinematics}
\label{kinematics2}
 Our sample was selected to present bi-symmetric H$\alpha$ EW features alignment with the gas kinematic major axis using MaNGA data. Our GMOS data show that similar behaviours are also observed in the central region as can be seen in Fig. \ref{fig:1-385124} and Figs. \ref{fig:1-296801}--\ref{fig:1-279073}. According to \citet{cheung16}, the large scale bi-symmetric H$\alpha$ EW features alignment with the major kinematic axis observed for Red Geysers can not be explained by gas in a disc. In that case, one would not expect the collimated by-symmetric emission structure being always oriented along the kinematic major axis of the galaxy. The observed second velocity moment  V$_{RMS}$= (V$^{2}$ + $\sigma^{2}$)$^{\frac{1}{2}}$ for the prototype Red Geyser -- the Akira galaxy -- is about 100 km s$^{-1}$ larger than the  predicted value for a rotating disc \citep{cheung16}.   \citet{riffel19}, using GMOS-IFU observations of Akira, found that the V$_{RMS}$ exceeds by up to $\sim$50 km s$^{-1}$ the V$_{RMS}$ predicted by \citet{cheung16} in the central region and outflows are needed to describe its kinematics. Furthermore, a bi-conical wind model reproduces the shape of the observed velocity field for Akira galaxy \citep{cheung16}. Our MaNGA and GMOS-IFU data reveal some perturbation in the velocity fields usually associated with higher velocity dispersion ($\sigma >150-200$ km s$^{-1}$) regions, which suggest signatures of ionised winds in these galaxies. The small and large scale velocity dispersion maps show $\sigma>200$ km s$^{-1}$ at the nucleus, being even larger than those seen for Akira.

 All the six Red Geysers with extended emission in our sample show some misalignment between the orientations of the gas kinematic major axis at small and large scales, as seen in Fig. \ref{fig:1-385124} and Figs. \ref{fig:1-296801}--\ref{fig:1-279073}, following the same behaviour observed in Akira  \citep{riffel19}, and interpreted as the variation in the outflow orientation being produced by the precession of the accretion disc caused by a misalignment between the spin of the black hole and the disc. Thus, our velocity maps indicate signatures of outflows changing the orientation from the nuclear region to kpc scales, as observed by \citet{riffel19} for Akira.

 \citet{roy21a} investigated the gas kinematics of two Red Geysers: Akira (MaNGA 1-217022) and 1-145922. They found asymmetric profiles for  the [N\,{\sc ii}] and H$\alpha$ emission lines with red wings on the blueshifted side of the velocity field and blue wings on the redshifted side. To explain the observed asymmetries, they proposed a wind model with bi-conical outflows. The high absolute values seen in the $h_{3}$maps based on the GMOS data observed for all galaxies in our sample indicate the presence of asymmetries in the emission line profiles, which are confirmed by visual inspection of the spectra.  Moreover, the $h_{3}>0$ (red wings) are co-spatial with the blueshifted locations of velocity fields, while the $h_{3}<0$ (blue wings) overlap with the redshifted regions, the same behaviour observed by \citet{roy21a}. Thus, the galaxies in our sample show clear signature of ionised gas winds, driven by a central AGN and likely produced by precession in the accretion disc, as found for Akira. A detailed analysis of the ionised gas kinematics, including the modelling and estimates of the outflow properties, will be presented in a forthcoming work.

\section{Conclusions}
\label{conclusions}
We have analysed the gas ionisation and kinematics in a sample of Red Geysers with IFS from MaNGA SDSS-IV and GMOS. The MaNGA-IFU observations have a spatial coverage ranging between 12\farcs0 and 32\farcs0, while the GMOS-IFU observations show a FoV of 5\farcs0$\times$3\farcs5. The MaNGA data have an angular resolution do 2\farcs5 corresponding to 1.3--3.7 kpc at the distance of galaxies. The GMOS-IFU has a spatial resolution of 0.3--0.9 kpc for our sample and spectral resolution of $\sim$ 1.6 \AA\ . The main conclusions we have reached are:
\begin{itemize}
\item The emission line ratios of all galaxies, within an aperture of 2\farcs5 in diameter (inner 1.3--3.7 kpc at the distance of galaxies) for MaNGA and nuclear apertures corresponding to the GMOS-IFU resolution (0.3--0.9 kpc at the distance of galaxies) indicate that the emission is produced by gas photoionised by Seyfert/LINER nuclei.

\item Only two galaxies, MaNGA 1-279073 and 1-296801, have H$\alpha$ equivalent width greater than 3 \AA\ with MaNGA data, which is a strong indication of the presence of an AGN. But the better spatial resolution of the GMOS-IFU data shows four Red Geysers with H$\alpha$ EW $>$ 3 \AA\: MaNGA 1-279073, 1-296801, 1-385124 and 1-197230.
\item Using  H$\alpha$ EW  $>$ 1.5 \AA\ as an indicator of AGN -- that we argue is better suited for faint AGN-- we find five Red Geysers in MaNGA and six in GMOS-IFU data. Five of them are in both data: MaNGA 1-279073, 1-296801, 1-385124, 1-197230 and 1-24104. The galaxy MaNGA 1-114245 has an AGN that is detected only in GMOS-IFU data.

\item  The Red Geysers MaNGA 1-24104, 1-279073 and 1-296801 are radio detected in the VLA-FIRST survey providing additional support that they are AGN hosts. 
\item Electron density measurements suggest a high-density gas ($N_{e} >$ 300 cm$^{-3}$) compared to the value previously determinate for the prototypical Red Geyser Akira galaxy ($N_{e} \sim$ 100 cm$^{-3}$). However, they are in agreement with the $N_{e}$ estimated for the NLR and outflow medium of AGNs. 
\item The large scale (MaNGA) and nuclear scale (GMOS) gas velocity fields are misaligned, with kinematic position angle differences between 12$^{\circ}$ and 60$^{\circ}$. The emission-line profiles are asymmetrical, with blue wings on the redshifted side of the velocity field and red wings on the blueshifted side.  These results support that Red Geysers host ionised gas outflows, originating in a precessing accretion disc.
\end{itemize}

\section*{Acknowledgements}
We thank the support of the Instituto Nacional de Ci\^encia
e Tecnologia (INCT) e-Universe (CNPq grant 465376/2014-2).
GSI acknowledges financial support from Conselho Nacional de Desenvolvimento Cient\'ifico e Tecnol\'ogico (CNPq -- 	142514/2018-7).
RAR acknowledges financial support from Conselho Nacional de Desenvolvimento Cient\'ifico e Tecnol\'ogico (CNPq -- 302280/2019-7) and Funda\c c\~ao de Amparo \`a pesquisa do Estado do Rio Grande do Sul (FAPERGS -- 21/2551-0002018-0). TVR thanks to CNPq (grant 306790/2019-0). RR thanks to CNPq (grant 311223/2020-6,  304927/2017-1 and  400352/2016-8) and FAPERGS (grant 16/2551-0000251-7 and 19/1750-2).

Funding for the Sloan Digital Sky Survey IV has been provided by the Alfred P. Sloan Foundation and the Participating Institutions.  SDSS-IV acknowledges
support and resources from the Center for High-Performance Computing at the University of Utah. The SDSS web site is www.sdss.org. SDSS-IV is managed by the
Astrophysical Research Consortium for the Participating Institutions of the SDSS Collaboration including the Brazilian Participation Group, the Carnegie Institution for Science,
Carnegie  Mellon  University,  the  Chilean  Participation  Group,  Harvard-Smithsonian Center for Astrophysics, Instituto de Astrof\'isica de Canarias, The Johns Hopkins University, Kavli Institute for the Physics and Mathematics of the Universe (IPMU) / University of Tokyo, Lawrence Berkeley National Laboratory, Leibniz Institut f\"ur Astro
physik  Potsdam  (AIP),  Max-Planck-Institut  f\"ur  Astrophysik  (MPA Garching),  Max-Planck-Institut f\"ur Extraterrestrische Physik (MPE), Max-Planck-Institut f\"ur Astronomie
(MPIA Heidelberg), National Astronomical Observatory of China, New Mexico State University, New York University, The Ohio State University, Pennsylvania State University, 
Shanghai  Astronomical  Observatory,  United  Kingdom  Participation  Group, Universidad Nacional Aut\'onoma de M\'exico, University of Arizona, University of Colorado Boulder,
University of Portsmouth, University of Utah, University of Washington, University of Wisconsin, Vanderbilt University, and Yale University.

\section*{Data Availability}

The data used in this paper are available in the Gemini Science Archive \url{https://archive.gemini.edu/searchform} under the codes of the projects GN-2020A-Q-226 and GN-2020A-Q-130. The processed data cubes used in this paper will be shared on reasonable request to the corresponding author.



\bibliographystyle{mnras}



\appendix

\section{GMOS sample maps}

\begin{figure*}
	\includegraphics[width=\textwidth]{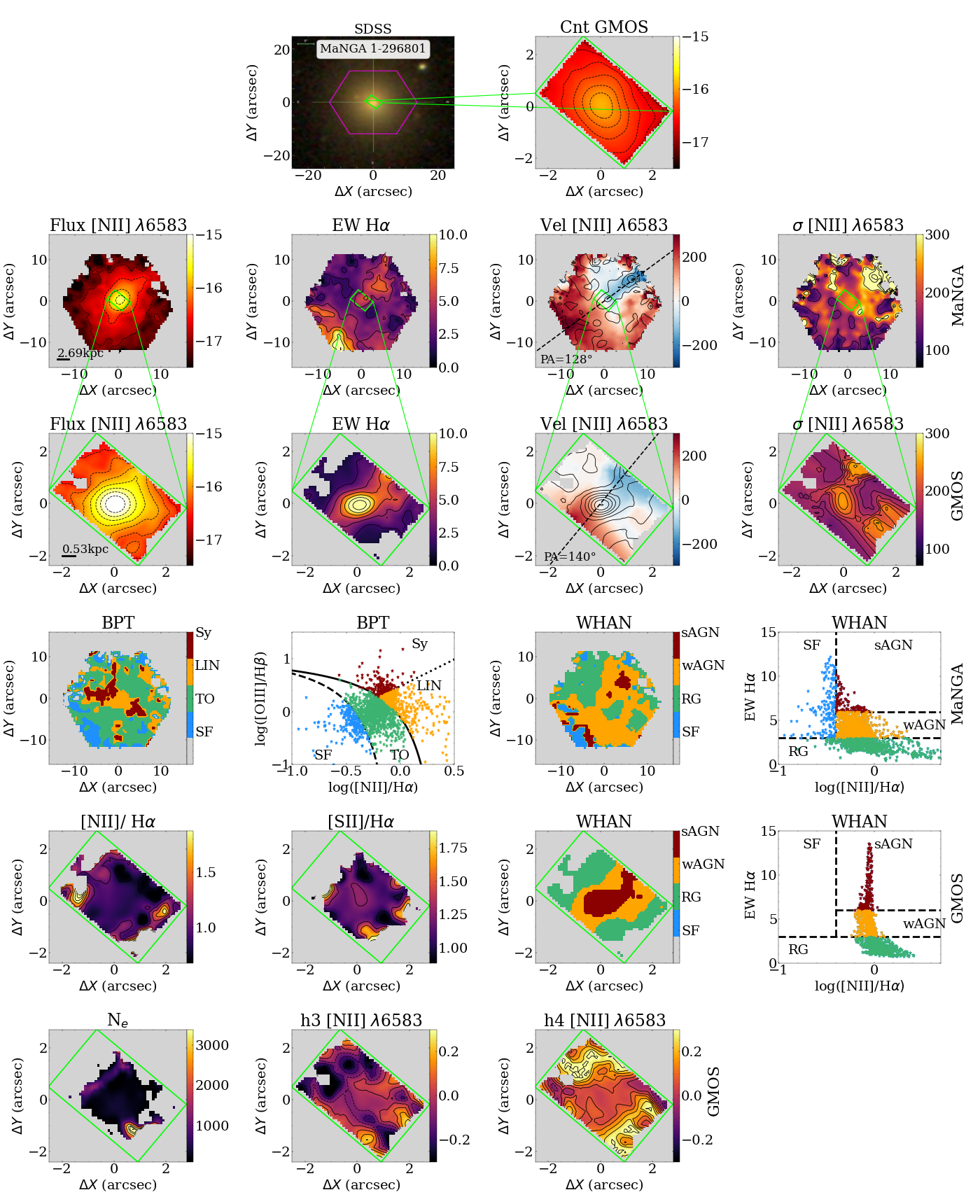}
  \caption{Same as Fig. \ref{fig:1-385124}, but for galaxy MaNGA 1-296801. }
    \label{fig:1-296801}
\end{figure*}

\begin{figure*}
	\includegraphics[width=\textwidth]{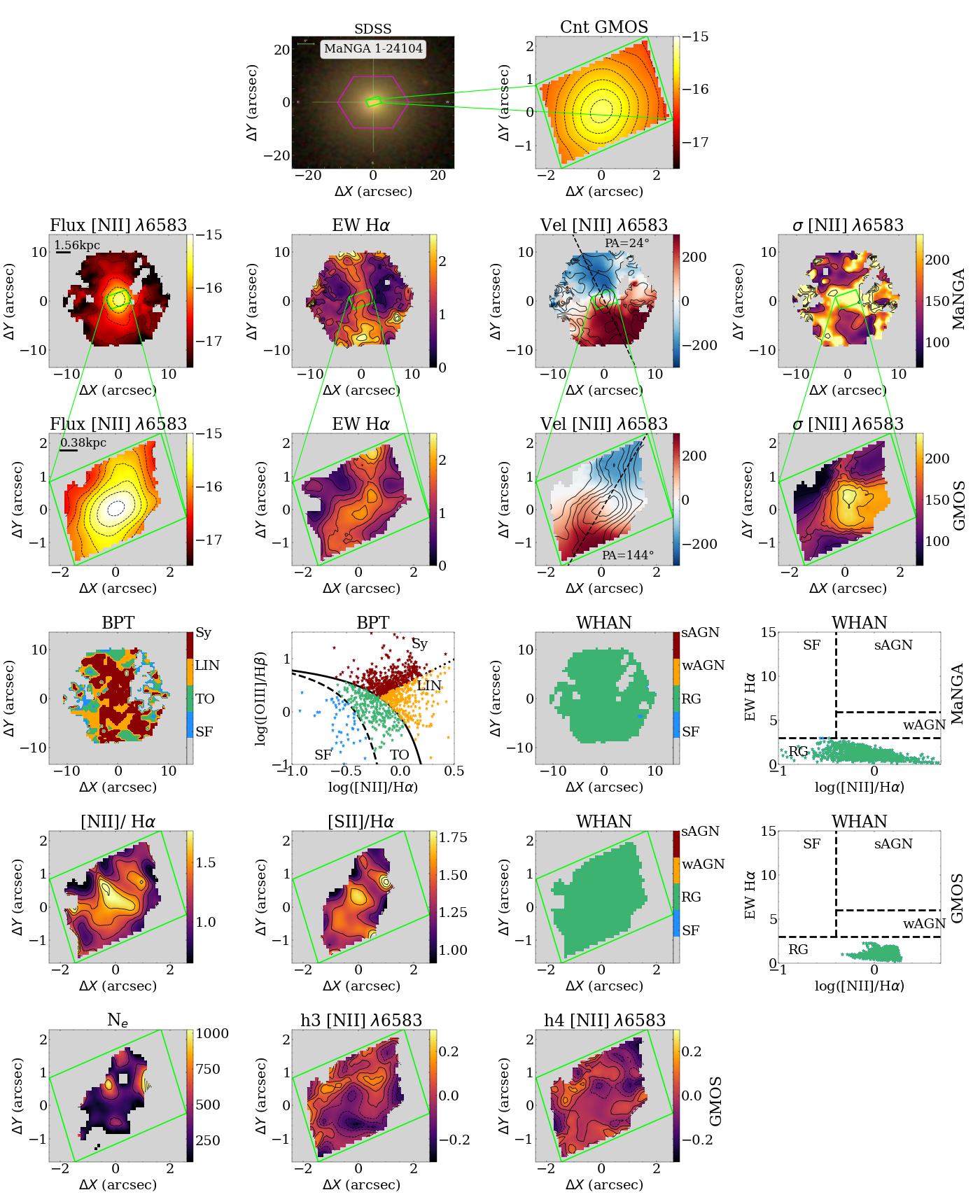}
    \caption{Same as Fig. \ref{fig:1-385124}, but for galaxy MaNGA 1-24104. }
    \label{fig:1-24104}
\end{figure*}

\begin{figure*}
	\includegraphics[width=\textwidth]{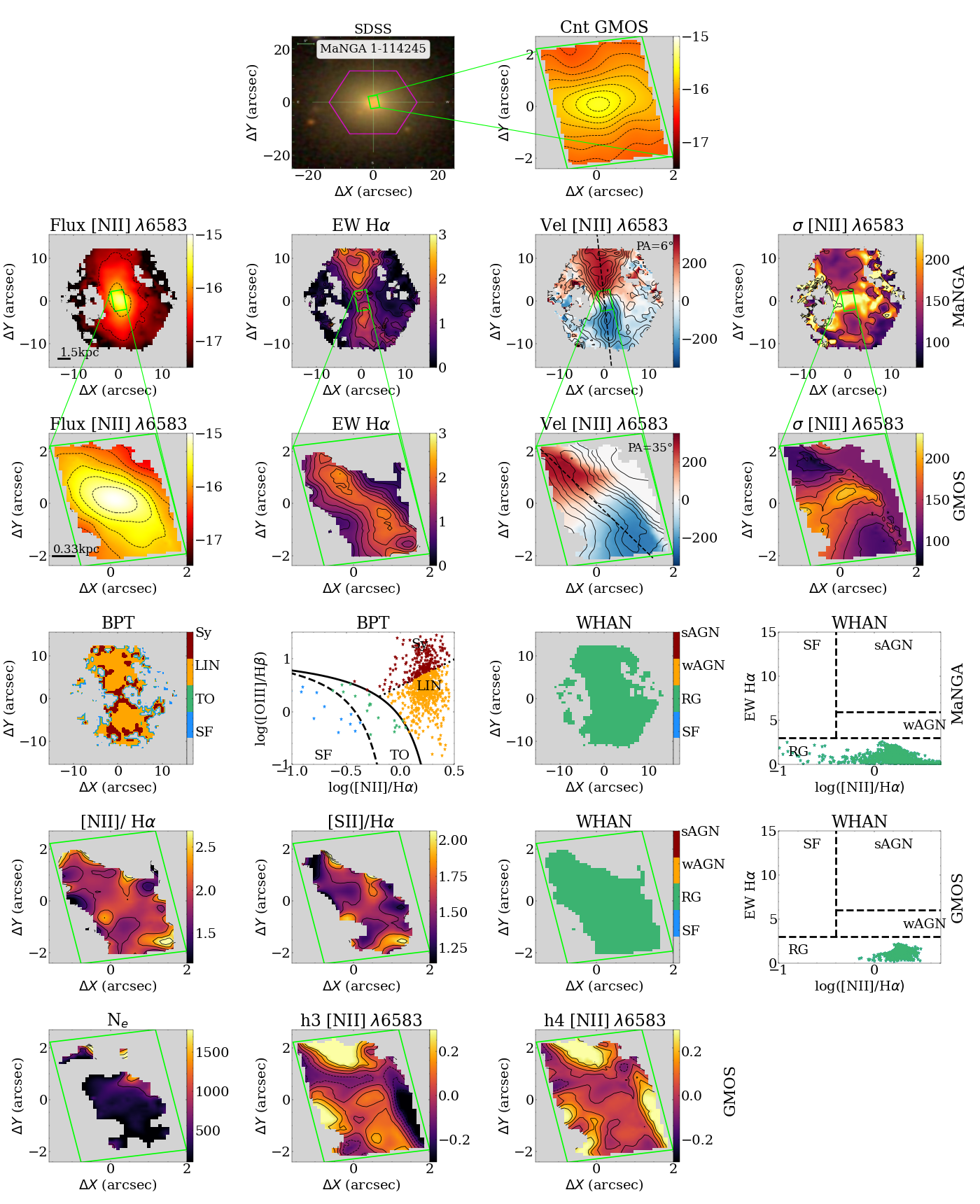}
    \caption{Same as Fig. \ref{fig:1-385124}, but for galaxy MaNGA 1-114245.}
    \label{fig:1-114245}
\end{figure*}
\begin{figure*}
	\includegraphics[width=\textwidth]{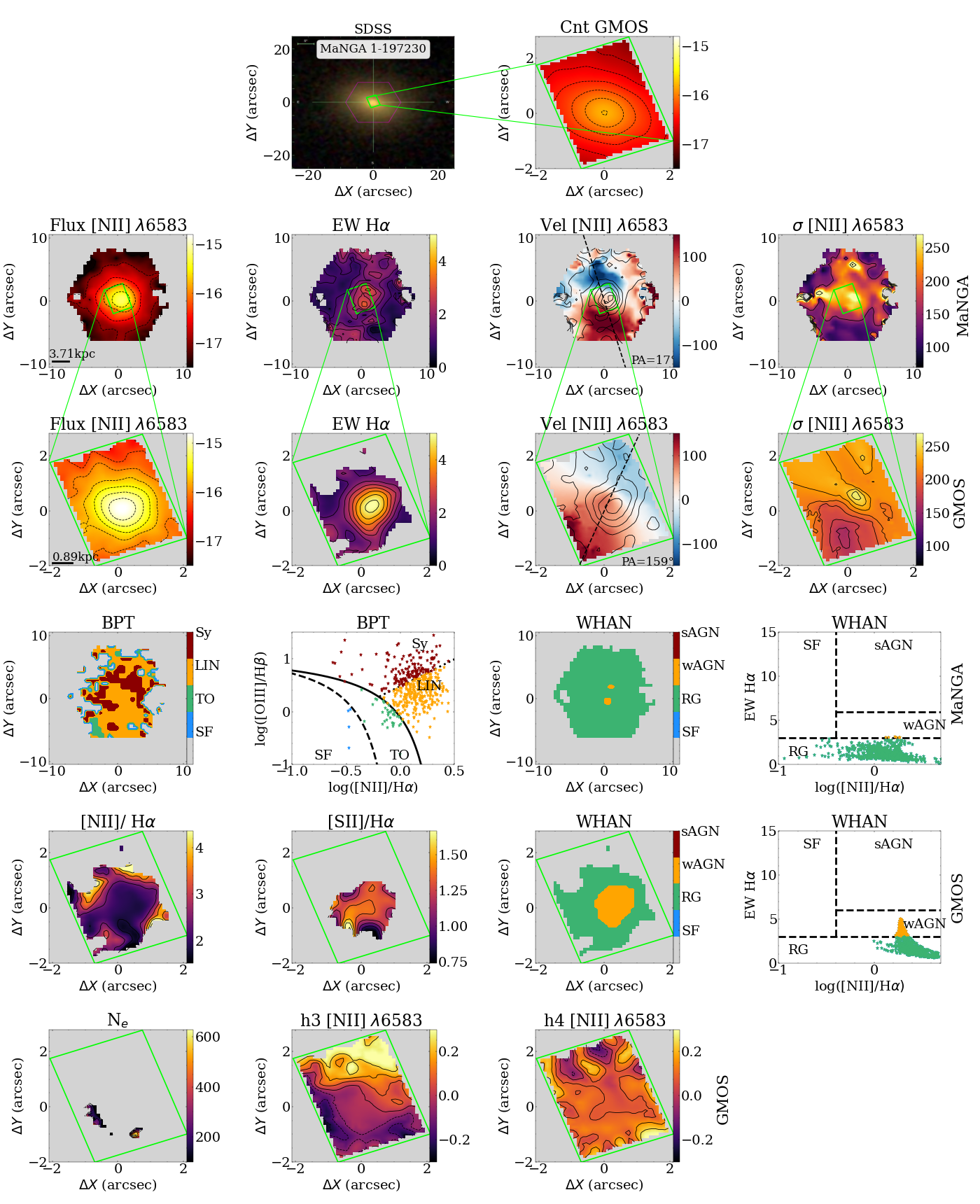}
    \caption{Same as Fig. \ref{fig:1-385124}, but for galaxy MaNGA 1-197230.}
    \label{fig:1-197230}
\end{figure*}

\begin{figure*}
	\includegraphics[width=\textwidth]{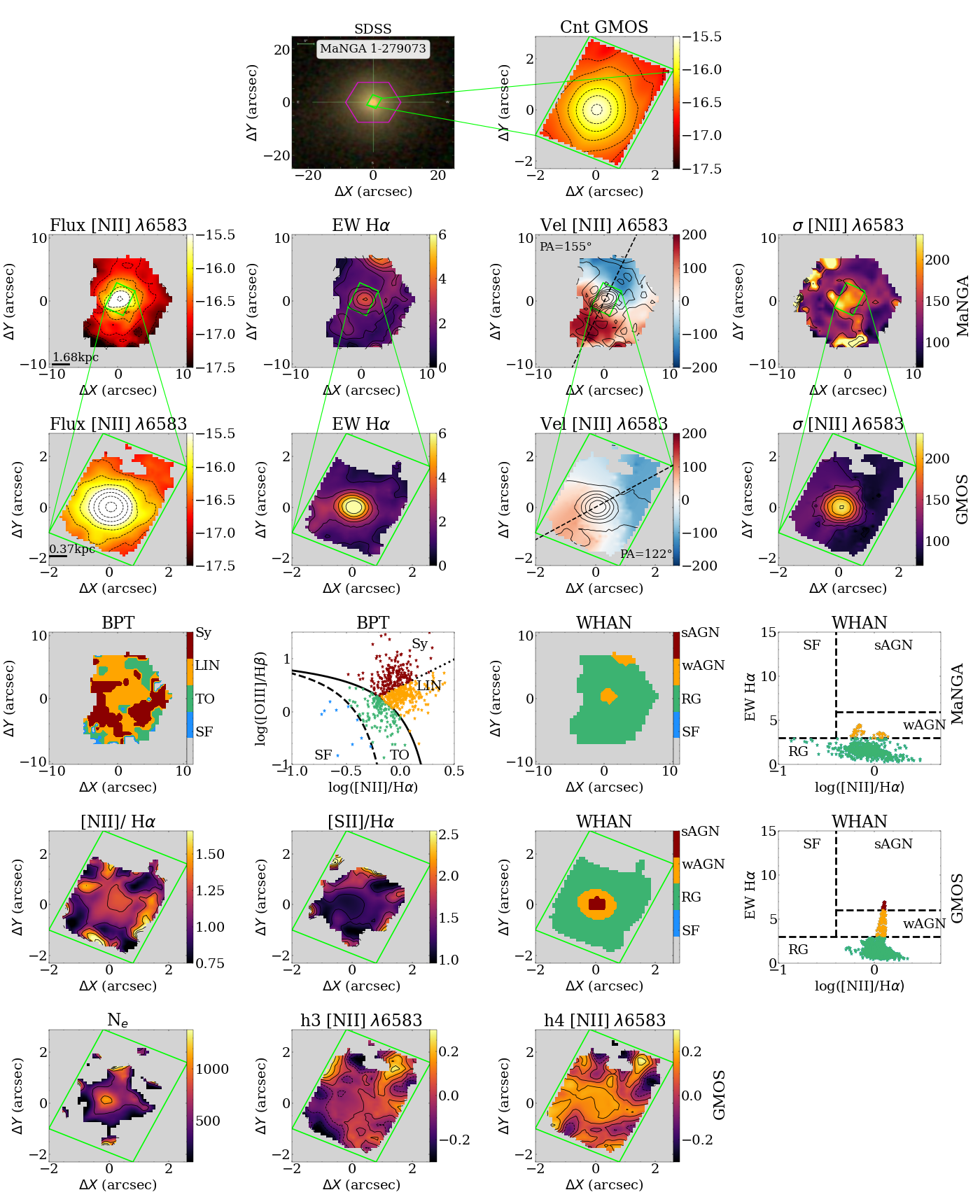}
    \caption{Same as Fig. \ref{fig:1-385124}, but for galaxy MaNGA 1-279073.}
    \label{fig:1-279073}
\end{figure*}

\begin{figure*}
	\includegraphics[width=0.50\textwidth]{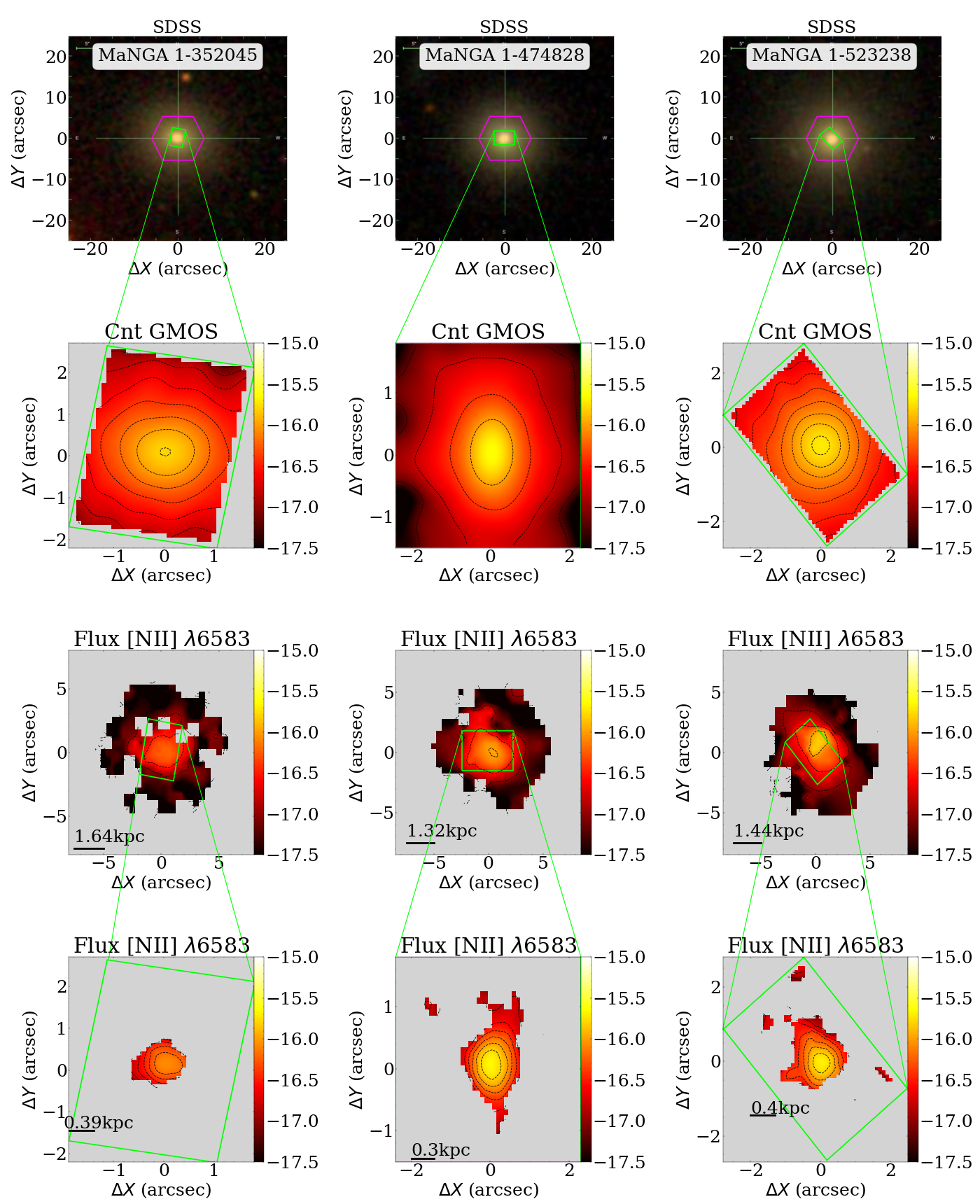}
    \caption{The maps were produced using MaNGA-DAP and modelling the emission-line profiles of GMOS-IFU data for three galaxies with compact emission. The first row presents the optical image from SDSS with the MaNGA-IFU in magenta. In the second row, the continuum maps from GMOS-IFU data are shown. The third and fourth rows show [N\,{\sc ii}]$\lambda$6583  flux distributions from MaNGA-DAP and GMOS-IFU data. The green rectangles indicate the GMOS-IFU FoV. The continuum and flux maps are in units of erg s$^{-1}$ cm$^{-2}$ arcsec$^{-2}$  \AA$^{-1}$ and  erg s$^{-1}$ cm$^{-2}$ arcsec$^{-2}$. }
    \label{fig:fluxt}
\end{figure*}

\begin{figure*}
	\includegraphics[width=0.75\textwidth]{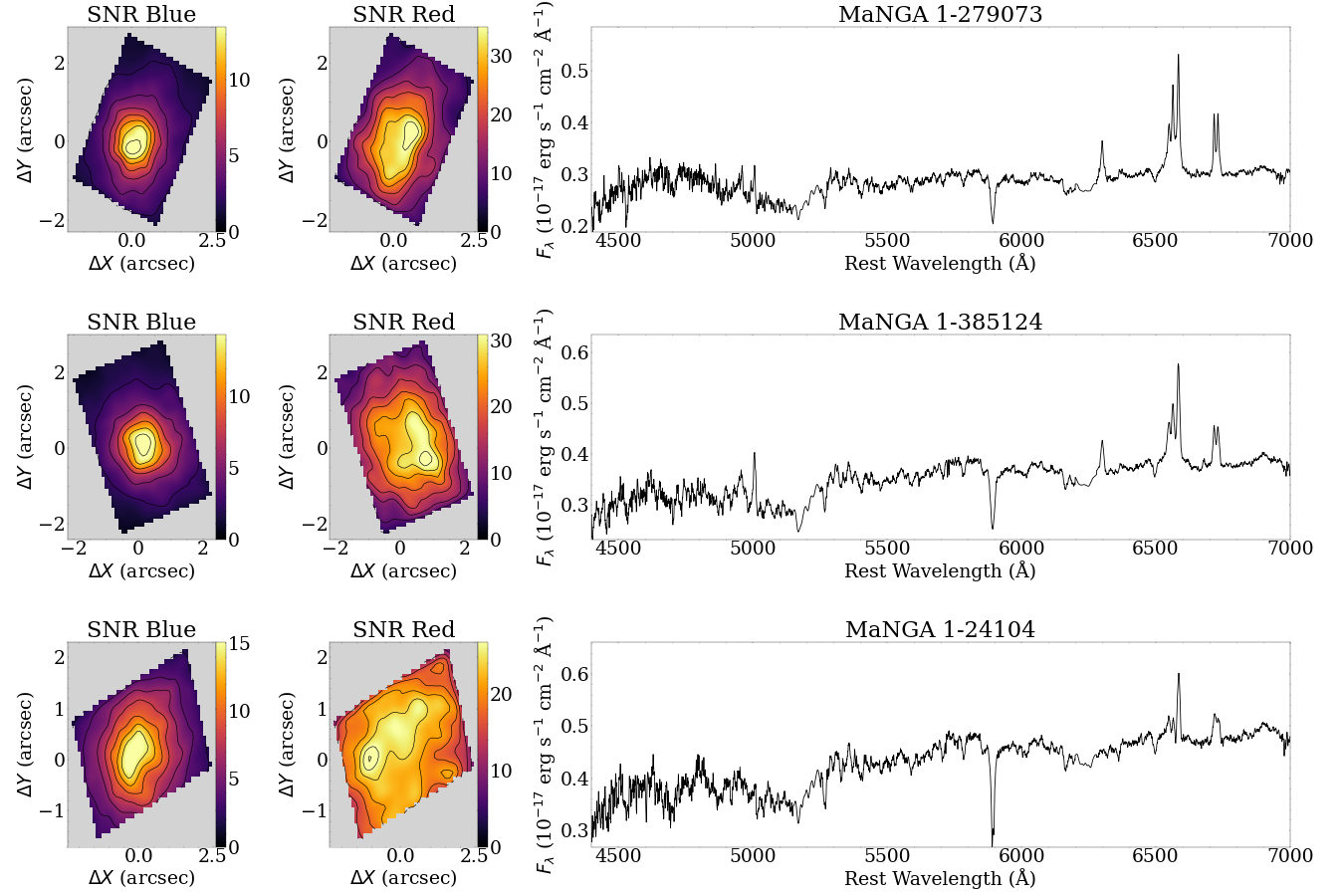}
    \caption{Signal-to-noise ratio estimated for continuum in a 300 \AA\ spectral window centred at 4650 \AA\ (SNR Blue), centred at 5650 \AA\ (SNR Red), and GMOS-IFU spectrum for the nuclear spaxel of the galaxies.}
    \label{fig:snr}
\end{figure*}

\begin{figure*}
	\includegraphics[width=0.75\textwidth]{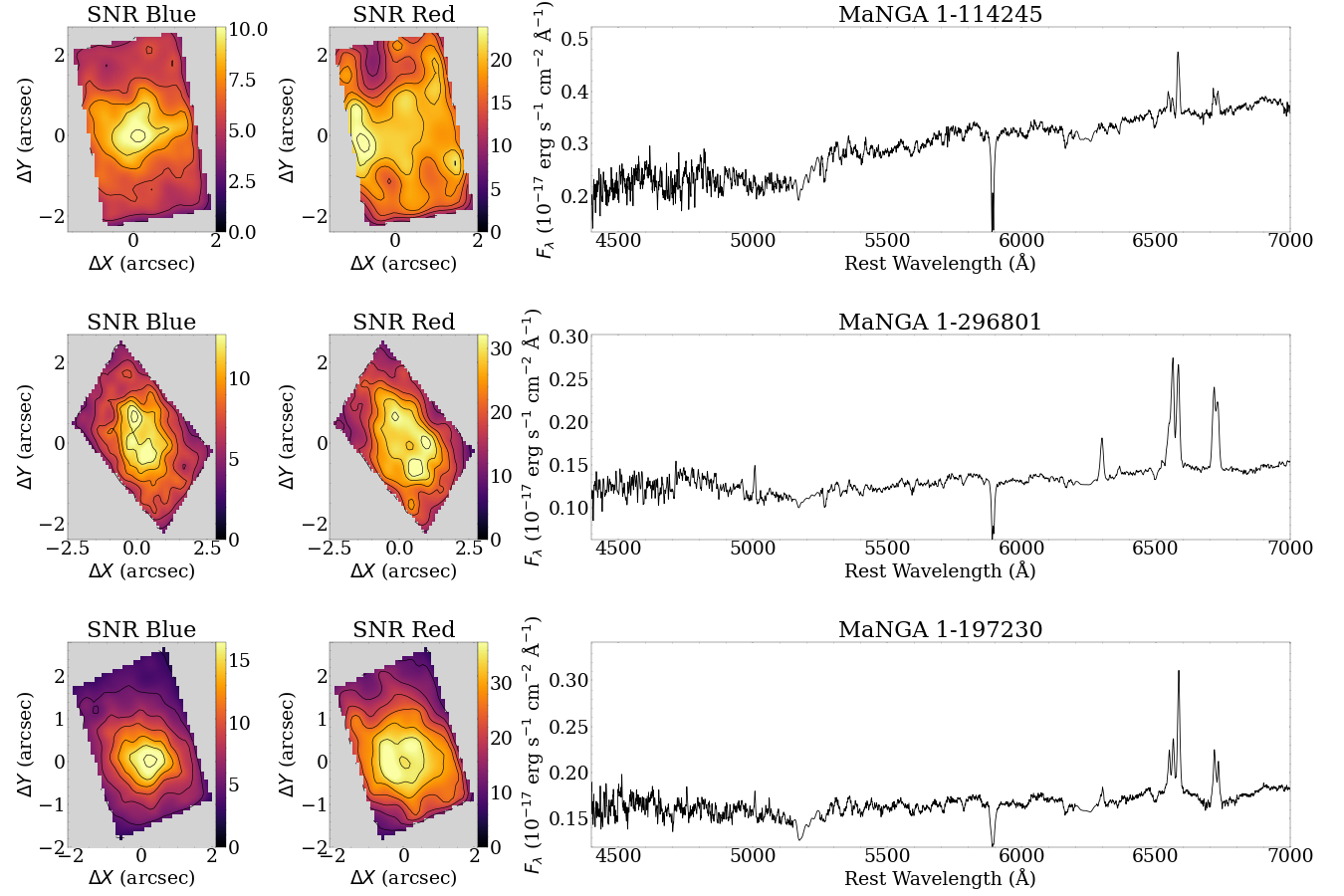}
    \caption{Same as Fig. \ref{fig:snr}.}
    \label{fig:snr1}
\end{figure*}

\begin{figure*}
	\includegraphics[width=0.75\textwidth]{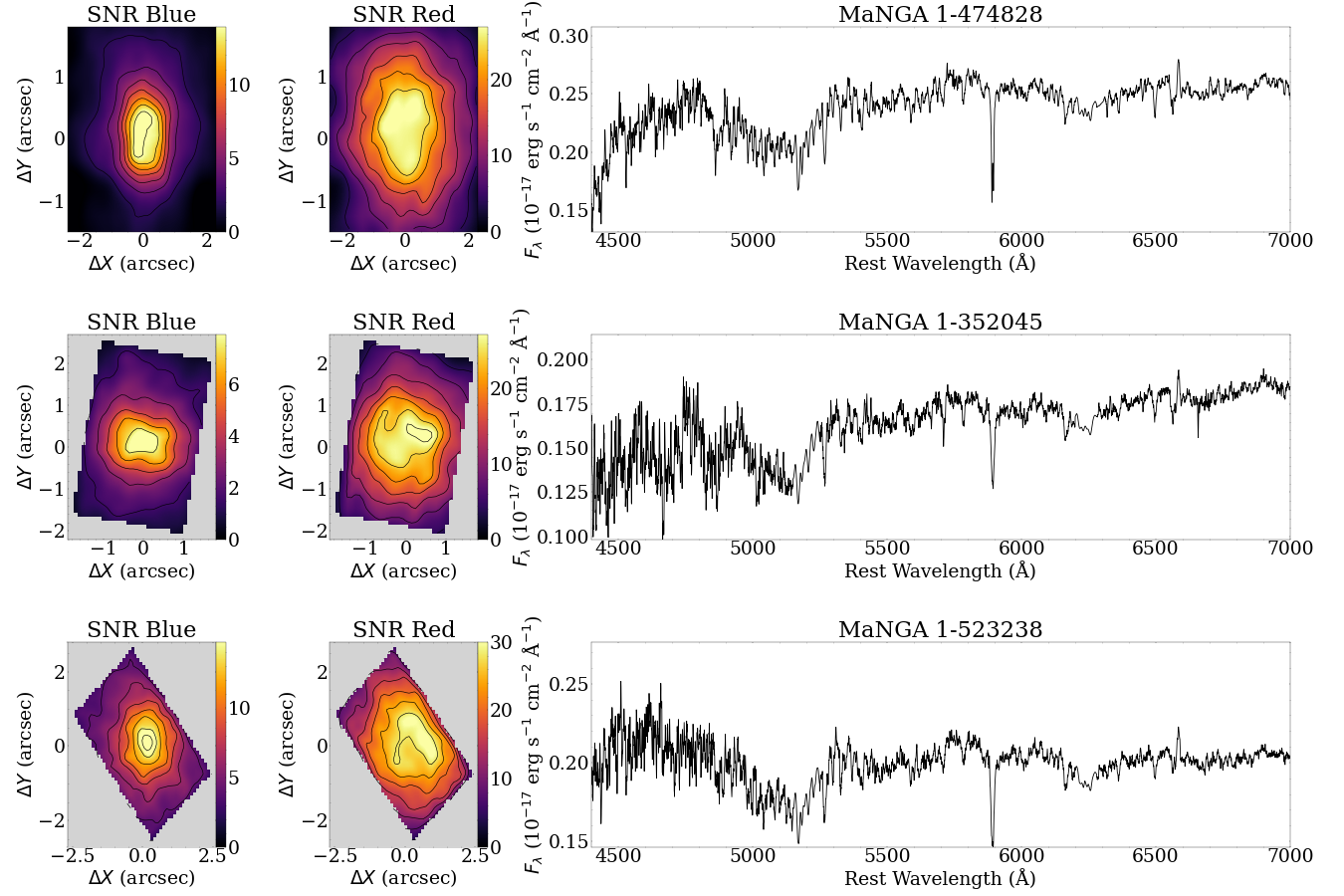}
    \caption{Same as Fig. \ref{fig:snr}.}
    \label{fig:snr2}
\end{figure*}


\bsp	
\label{lastpage}
\end{document}